\documentclass[11pt ]{article}
\usepackage{amsmath}
\usepackage{graphicx}
\usepackage{hyperref}
\usepackage{lmodern}
\usepackage{amssymb}
\usepackage{booktabs}
\usepackage{tikz}
\usetikzlibrary{patterns}
\renewcommand{\theequation}
{\arabic{section}.\arabic{equation}}

\makeatletter
\def\eqnarray{ \stepcounter{equation} \let\@currentlabel=\theequation
 \global\@eqnswtrue
 \global\@eqcnt\z@
 \tabskip\@centering
 \let\\=\@eqncr
 $$\halign to \displaywidth\bgroup\@eqnsel\hskip\@centering
 $\displaystyle\tabskip\z@{##}$&\global\@eqcnt\@ne
 \hfil$\displaystyle{{}##{}}$\hfil
 &\global\@eqcnt\tw@$\displaystyle\tabskip\z@{##}$\hfil
 \tabskip\@centering&\llap{##}\tabskip\z@\cr}
\makeatother

\makeatletter
\def\@arrayacol{\edef\@preamble{\@preamble \hskip .5\arraycolsep}}
\def\array{\let\@acol\@arrayacol \let\@classz\@arrayclassz
\let\@classiv\@arrayclassiv \let\\\@arraycr\def\@halignto{}\@tabarray}
\makeatother

\makeatletter
\newcounter{subeqncnt}
\def\thesubeqncnt{\alph{subeqncnt}}
\def\subequations{\begingroup%
   \stepcounter{equation}\edef\@tempa{\theequation}%
   \let\c@equation\c@subeqncnt\c@subeqncnt\z@
   \edef\theequation{\@tempa\noexpand\thesubeqncnt}}

\makeatother

\newcommand{\be}{\begin{equation}}
\newcommand{\ee}{\end{equation}}

\newcommand{\beqa}{\begin{eqnarray}}
\newcommand{\eeqa}{\end{eqnarray}}
\newcommand{\nn}{\nonumber}

\def\CB {{\cal B}}

\def\CD {{\cal D}}
\def\CE {{\cal E}}
\def\CF {{\cal F}}
\def\CG {{\cal G}}
\def\CH {{\cal H}}

\def\CL {{\cal L}}
\def\CM {{\cal M}}

\def\CO {{\cal O}}

\begin{document}

\setlength{\baselineskip}{7mm}
\begin{titlepage}
\begin{flushright}
{\tt NRCPS-HE-77-2019} \\
June, 2019
\end{flushright}

\vspace{1cm}

\begin{center}
{\it \Large From Heisenberg-Euler Lagrangian \\ to the discovery \\
of Chromomagnetic Gluon Condensation\footnote{
Based on lectures at the Leipzig  University in occasion of the 80 Years of Heisenberg-Euler Lagrangian 1936-2016, ITP, Leipzig, November 21, 2016  and
40 Years of Discovery of the Chromomagnetic Gluon Condensation 1977-2017 at the Ludwig-Maximilian University München,
Arnold Sommerfeld Colloquium at Center for Theoretical Physics,  April 18, 2018.}  
}

\vspace{1cm}

{ {G.~K.~Savvidy  }}

\vspace{1cm}

 {\it Institute of Nuclear and Particle Physics,} \\
{\it Demokritos National Research Center }\\
{\it Agia Paraskevi, GR-15310 Athens, Greece}

\end{center}

\vspace{1cm}

\begin{abstract}
I reexamine the phenomena of the chromomagnetic gluon condensation in Yang-Mills theory. The extension of the Heisenberg-Euler Lagrangian to the Yang-Mills theory allows to calculate the effective action, the energy-momentum tensor and demonstrate that the energy density curve crosses the zero energy level of the perturbative vacuum state at nonzero angle and continuously enters to the negative energy density region.  At the crossing point and further down  the effective coupling constant is small and demonstrate that the true vacuum state of the Yang-Mills theory is below the perturbative  vacuum state and is described by the nonzero chromomagnetic gluon condensate. The renormalisation group analysis allows to express the energy momentum tensor, its trace  and the first and second order derivatives  in terms of Callan-Symanzik beta function and effective coupling constant. The derivatives define the convexity  and the extremum of the energy density curve. In the vacuum the energy-momentum tensor is proportional to the space-time metric, and induces a negative contribution to the effective cosmological constant.  

\end{abstract}

\end{titlepage}

\section{\it Introduction}

In this article we shall analyse the effective action in QED  and QCD by using the perturbative loop expansion and renormalisation group equations and discuss the physical consequences which can be derived from their explicit expressions.  We shall reexamine the phenomena of the chromomagnetic gluon condensation in Yang-Mills (YM) theory and will present the derivation of the new results. The Heisenberg-Euler Lagrangian in QED \cite{Sauter:1931zz,Heisenberg:1934pza,Euler:1935zz,Heisenberg:1935qt,Schwinger:1951nm,Coleman:1973jx} is a sum of the one loop diagrams with a vacuum electron-positron pair circulating in the loop and the gluons and quarks in case of QCD  \cite{Vanyashin:1965ple,Skalozub:1975ab,Brown:1975bc,Duff:1975ue,PhDTheses,Batalin:1976uv,Savvidy:1977as, Matinyan:1976mp}.  The effective action $\Gamma[A]$ has the following representation:  
\beqa\label{effectiveaction}
 \Gamma =\int \CL dx &=& \sum_n \int dx_1...dx_n \Gamma^{(n) a_1...a_n}_{~~~\mu_1...\mu_n}(x_1,...,x_n) A^{a_1}_{\mu_1}(x_1)...
 A^{a_n}_{\mu_n}(x_n)  = S + W^{(1)} + W^{(2)}+..,~~~~~~~~~~
 \eeqa 
where $\CL$ is the effective Lagrangian, $ \Gamma^{(n)}$ is a one-particle irreducible (1PI) n-point vertex function,  $ A^{a}_{\mu}(x) \equiv  \langle 0\vert A^{a}_{\mu}(x) \vert 0 \rangle$ is the vacuum expectation value of the field operator and  $W^{(n)},~n=1,2,..$ represent the terms of the loop expansion.

\begin{figure}
 \centering
\includegraphics[angle=2.5,width=8cm]{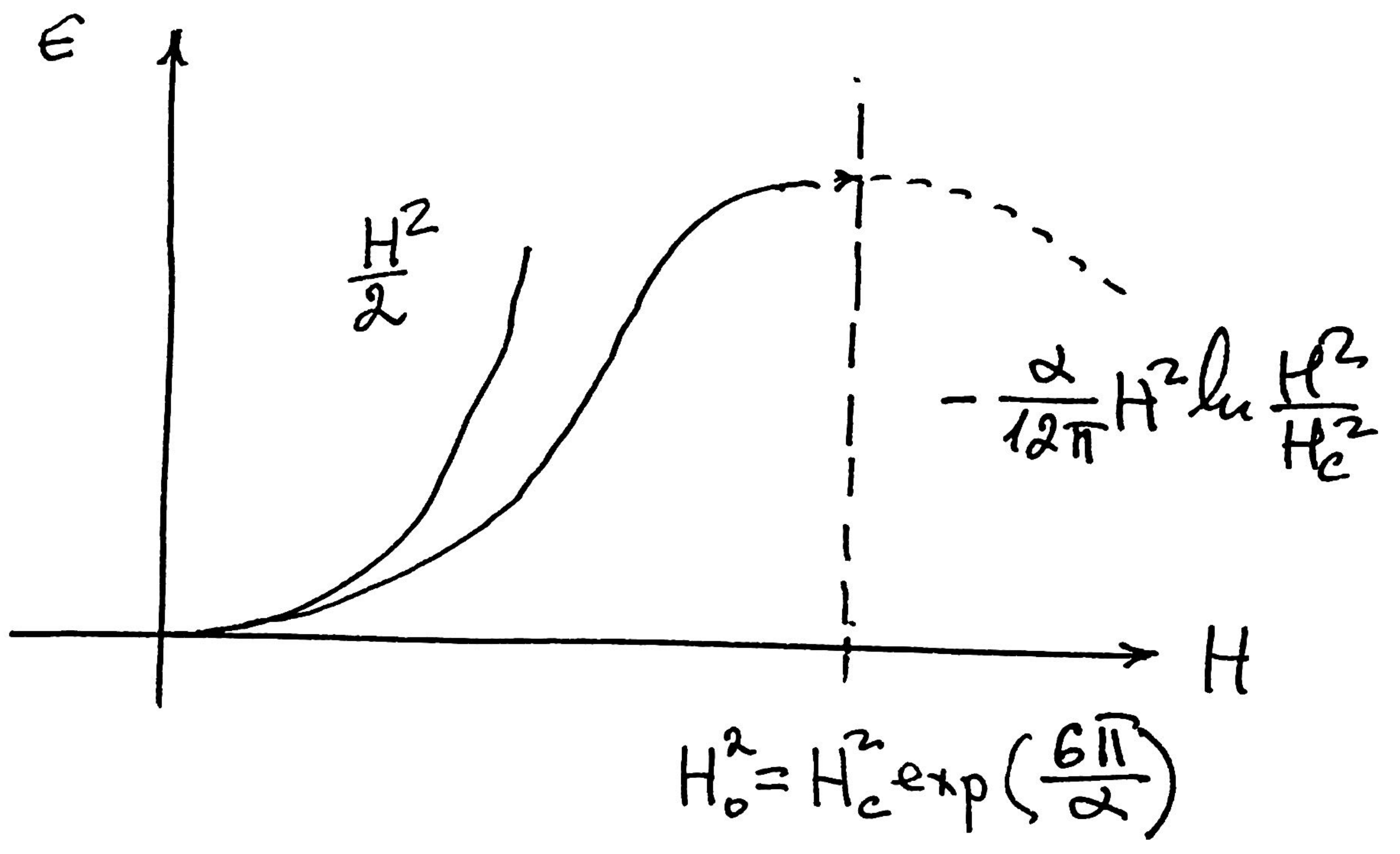}
\centering
\caption{The graph shows the qualitative behaviour of the QED vacuum energy density $\epsilon(\vec{\CH}^2)$ (\ref{QED0}),(\ref{energyexp}) and of the effective coupling constant $\bar{e}^2(\vec{\CH}^2)$ (\ref{effectivee})  as the  functions of the background magnetic field.  The effective coupling constant  is singular  at $ \vec{\CH_0}^2$,  the "Moscow zero" \cite{landau1,ioffe}.}
\label{fig1} 
\end{figure}

\begin{figure}
 \centering
\includegraphics[angle=0,width=8cm]{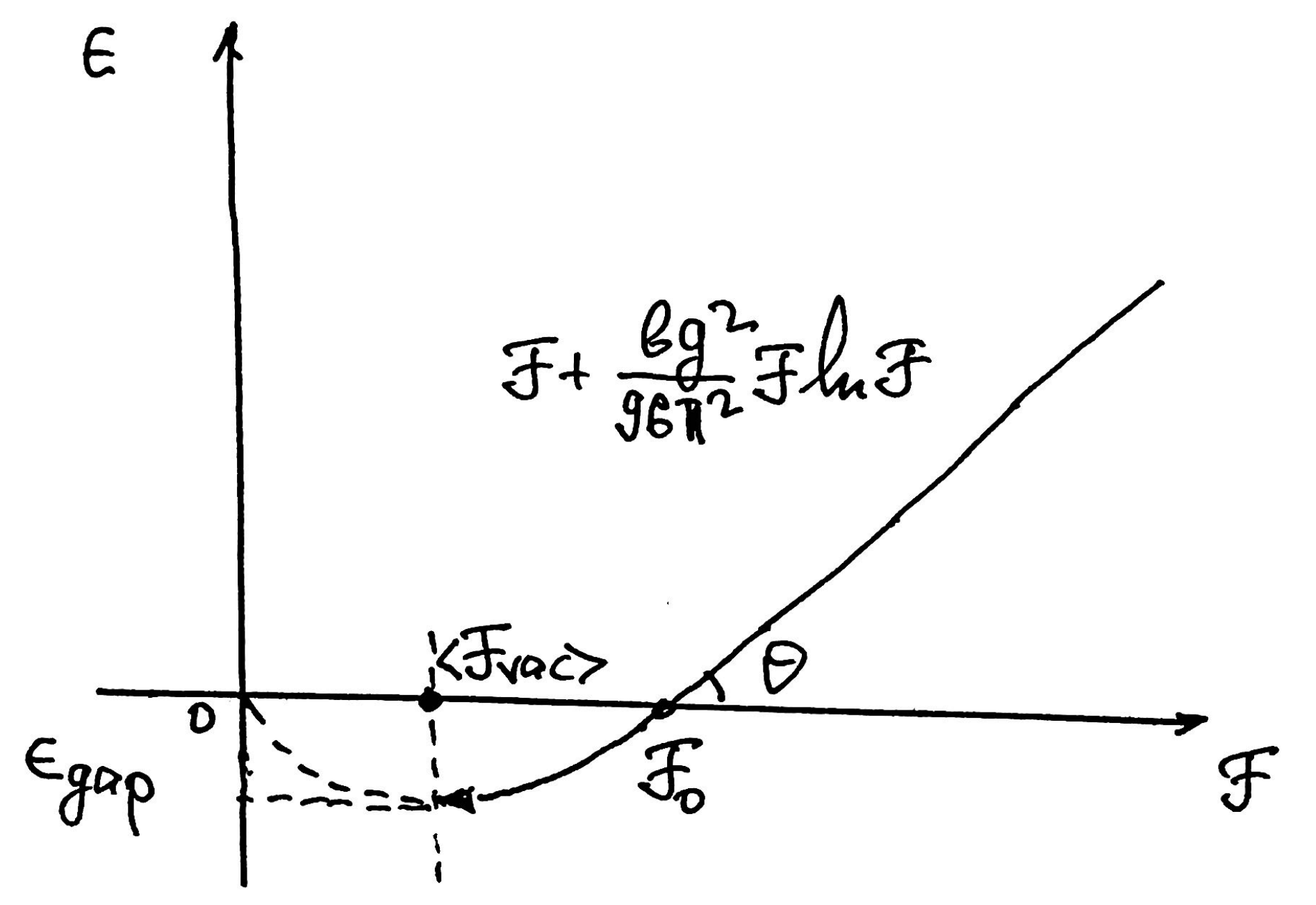}
\centering
\caption{The graph shows the qualitative behaviour of the vacuum energy density $\epsilon(\CF)$ (\ref{energyexpYM0}), (\ref{vacuumenergy}). At the intersection point $\CF_0$ (\ref{intersection}) the effective coupling constant is  small   (\ref{effcoupling}) and the intersection angle $\theta $ is strictly positive (\ref{intersectionangle}).  The energy density curve can be continuously extended from the point $\CF_0$  deep into the negative energy density region arbitrary close to the value of the vacuum condensate $\langle  \CF \rangle_{vac}$ by considering a larger values of $N$ and keeping  the t'Hooft coupling constant $g^2 N$ fixed (\ref{infraredpole}), (\ref{infraredpole1}).  This demonstrate that  there is a nonzero energy gap $\epsilon_{gap} > 0$ between perturbative and true vacuum states.  The vacuum is characterised by the nonzero value of the chromomagnetic field strength tensor (\ref{chomomagneticcondensate0}), (\ref{chomomagneticcondensate}) and  the energy density  gap $\epsilon_{gap} = \vert \epsilon_{vac} \vert $ (\ref{gap}) \cite{Savvidy:1977as,Nielsen:1978zg}.
}
\label{fig2} 
\end{figure}

We shall consider the limit of massless electrons  and quarks and demonstrate that the proper time integral in the Heisenberg-Euler Lagrangian can be calculated explicitly by using covariant renormalisation condition  \cite{PhDTheses,Savvidy:1977as, Matinyan:1976mp}
 \be\label{renormcondition1}
 {\partial  \CL \over \partial \CF} \vert_{t = {1\over 2}\ln ({2e^2  \vert \CF \vert \over \mu^4})= \CG=0} =-1, 
 \ee
where $\CF = {1\over 4}G^a_{\mu\nu}G^a_{\mu\nu}$ is the Lorentz and gauge invariant form of the YM field strength tensor $G^a_{\mu\nu}$ and $\mu^2$ is the renormalisation scale parameter.  In the massless limit the  QED effective Lagrangian has the exact logarithmic dependence as a function of  the invariant $\CF$ (see Fig.\ref{fig1}):
\beqa\label{QED0}
 \CL_e
&=& -\CF + {   e^2 \CF  \over 24 \pi^2} 
\Big[  \ln ({2 e^2 \CF \over \mu^4})  - 1  \Big] ,~~~~~~~~~ \CF =   {\vec{\CH}^2 - \vec{\CE}^2 \over 2},~~~
\CG =  \vec{\CE} \vec{\CH} =0,
\eeqa
where $\vec{\CH}$ and $\vec{\CE}$ are magnetic and electric fields. This expression should be compared with the one-loop effective Lagrangian in pure SU(N) gauge field theory, which has the form \cite{PhDTheses,Savvidy:1977as} (see Fig.\ref{fig2}):
\be\label{YMeffective0}
\CL_g  =  
-\CF - {11  N \over 96 \pi^2} g^2 \CF \Big( \ln {2 g^2 \CF \over \mu^4}- 1\Big)~,~~~~~~~~~ \CF =   {\vec{\CH}^2_a - \vec{\CE}^2_a \over 2} >0,~~~
\CG =  \vec{\CE}_a \vec{\CH}_a =0~.
\ee
From (\ref{QED0}) it follows that the corresponding quark contribution considered in the  chiral limit is
\beqa\label{chirallimit}
 \CL_q&=& -\CF + {    N_f  \over 48 \pi^2}  g^2 \CF
\Big[  \ln ({2 g^2 \CF \over \mu^4})  - 1  \Big] ~,~
\eeqa
where $N_f$ is the number of quark flavours.  

The effective Lagrangian technique allows to calculate the magnetic induction $\vec{\CB}$ of the vacuum  defined through the derivative of the effective Lagrangian \cite{PhDTheses}:
\be\label{palarYM0}
\vec{\CB}_a = - {\partial \CL \over \partial \vec{\CH}_a} ~=~ \mu_{vac}~   \vec{\CH}_a.
\ee
From (\ref{QED0}), (\ref{YMeffective0}) and (\ref{chirallimit}) it follows that in QED the vacuum responds to the background magnetic field as diamagnet and in QCD as paramagnet  with the magnetic permeabilities of the following form \cite{PhDTheses}:
\beqa\label{permeabilityQED0}
\mu_{QED} &=& 1- {e^2  \over 24 \pi^2}\log({e^2 \vec{\CH}^2 \over \mu^4}) ~< 1,~~~~~~~~~~~~~~~~~~~\text{\it diamagnetic},
\\ \label{permeabilityQCD0}
\mu_{QCD} &=& 1+ {  g^2  \over 96 \pi^2}(11N    - 2  N_f) \log{g^2 \vec{\CH}_a^2 \over \mu^4} ~ >~ 1,~~~~~\text{\it paramagnetic}, ~~~~N > {2\over 11} N_f.
\eeqa
The diamagnetism of the QED vacuum (\ref{permeabilityQED0}) means that it repels the magnetic fields by forming induced magnetic field in the direction opposite to that of the applied magnetic field. This phenomenon is similar to  the Landau orbital diamagnetism of free electron gas when the counteracting field is formed when the electron trajectories are curved due to the Lorentz force \cite{landau}. The paramagnetism of the QCD vacuum (\ref{permeabilityQCD0}) means that it amplifies  the applied chromomagnetic field by generating  induced chromomagnetic field in the direction of the applied field. In QCD the large polarisation of the gluon spins is responsible for the amplification of the background field.  This phenomenon is similar to the Pauli paramagnetism, an effect associated with the polarisation of the electron spins \cite{pauli}. 

The effective Lagrangian approach allows to calculate the quantum-mechanical corrections to the energy momentum tensor by using the formula derived by Schwinger in \cite{Schwinger:1951nm}:
\beqa\label{tmunu}
T_{\mu\nu} &=& (F_{\mu\lambda} F_{\nu\lambda}-g_{\mu\nu} {1\over 4} 
F^2_{\lambda\rho} ) {\partial \CL \over \partial \CF}- g_{\mu\nu} (\CL -
\CF {\partial \CL \over \partial \CF} - \CG {\partial \CL \over \partial \CG}). 
\eeqa
In case of the Heisenberg-Euler effective Lagrangian Schwinger  presented   the expression for the $T_{\mu\nu}$  in the fine structure constant $\alpha = e^2/4\pi $ expansion: 
\beqa 
T_{\mu\nu} &=& T^{M}_{\mu\nu}\Big( 1- {16  \over 45 m^4} \alpha^2 \CF \Big) + 
g_{\mu\nu}  {2  \over 45 m^4} \alpha^2 \Big( 4 \CF^2 + 7 \CG^2 \Big)  +... 
\eeqa
with its nonzero trace
\beqa
T &=& T_{\mu\mu} =  {8  \over 45 m^4} \alpha^2 \Big( 4 \CF^2 + 7 \CG^2 \Big) +...
\eeqa
In massless QED using the one-loop expression (\ref{QED0}) for $T_{\mu\nu}$ one can get 
\beqa\label{EnergyMomenQED0}
T_{\mu\nu} &=& T^{M}_{\mu\nu}\Big[1 - {   e^2   \over 24 \pi^2} 
  \ln {2 e^2 \CF \over \mu^4}   \Big]
+ g_{\mu\nu}    {   e^2   \over 24 \pi^2} \CF  ,~~~~~~~~~~~~\CG=0.
\eeqa
The $T_{\mu\nu}$  becomes proportional to the space-time metric tensor $g_{\mu\nu}$ at the extreme magnetic field $H^2_0 = H^2_c \exp{(6 \pi / \alpha)}$  and therefore induces  a positive effective cosmological constant  (see   Fig.\ref{fig1}).

To calculate the energy momentum tensor $T_{\mu\nu}$ in pure $SU(N)$ YM theory one should use the  expression (\ref{YMeffective0}) and  in the case of QCD,  in the limit of chiral fermions,  one should also add the quark contribution (\ref{chirallimit}) by using the substitution $11N \rightarrow b =11N - 2N_f$: 
\beqa\label{energymomentumYM0}
T_{\mu\nu} = T^{YM}_{\mu\nu}\Big[1 +{ b \ g^2 \over 96 \pi^2} 
  \ln {2 g^2 \CF \over \mu^4} \Big]
- g_{\mu\nu}    { b \   g^2 \over 96 \pi^2}  \CF  ,~~~~~~~~~\CG=0.
\eeqa
The vacuum energy density $T_{00} \equiv  \epsilon(\CF)$ has therefore the following form \cite{Savvidy:1977as}:
\be\label{energyexpYM0}
\epsilon(\CF)=  ~\CF + {b\ g^2\over 96 \pi^2}  \CF \Big( \ln {2 g^2 \CF \over \mu^4}- 1\Big).
\ee
The energy density has its new minimum outside of the perturbative vacuum  state $ \langle G_{\mu\nu}^2\rangle =0$, at the Lorentz and renormalisation group invariant field strength \cite{Savvidy:1977as}
\be\label{chomomagneticcondensate0}
\langle  2 g^2 \CF \rangle_{vac}=    \mu^4  \exp{(-{96 \pi^2 \over b\ g^2(\mu) })}= \Lambda^4_{QCD}, 
\ee
where $b = 11N -2N_f$ and characterises the dynamical breaking of  scaling invariance  in  YM theory\footnote{The $\Lambda_{QCD}$ is defined here through the covariant subtraction scheme (\ref{renormcondition1}). The relation with other renormalisation schemes was derived  in \cite{Nielsen:1978zg}.}:
\be
T_{\mu\mu}=  - { b    \over 48 \pi^2}  \langle 2 g^2 \CF \rangle_{vac} .\nn
\ee 
Substituting the vacuum field intensity     (\ref{chomomagneticcondensate0}) into the expression for the energy momentum tensor (\ref{energymomentumYM0}) one can get that in the vacuum the tensor $T_{\mu\nu}$ is proportional to the space-time metric  $g_{\mu\nu} $:
\be\label{vacuumenergyden0}
\langle T_{\mu\nu} \rangle_{vac} = - g_{\mu\nu}   ~ { b    \over 96 \pi^2} \langle g^2 \CF \rangle_{vac}. 
\ee
In this form the energy momentum tensor represents the relativistically invariant equation of state $\epsilon_{vac} =-P_{vac}$, which uniquely  characterises the vacuum \cite{Zeldovich:1968ehl,Weinberg} with its negative energy density $\epsilon_{vac} $.  The vacuum energy momentum tensor (\ref{vacuumenergyden0}) generates the effective cosmological constant $ \Lambda_{eff}$  
$$
R_{\mu\nu} - {1\over 2} g_{\mu\nu} R = g_{\mu\nu}   \Lambda_{eff} + {8\pi G \over c^4}   T_{\mu\nu} 
= {8\pi G \over c^4}   ( \langle T_{\mu\nu} \rangle_{vac} +   T_{\mu\nu} )
$$
of the form:
\be\label{gap}
\epsilon_{vac}  = {c^4   \Lambda_{eff} \over 8 \pi G  } ~=-{ b  \over 96 \pi^2}  \langle  g^2 \CF \rangle_{vac}= -{ b   \over 192 \pi^2} ~\Lambda^4_{QCD}  ~,
\ee
where the chromomagnetic condensate (\ref{chomomagneticcondensate0}) is   $< 2 g^2 \CF>_{vac}=\Lambda^4_{QCD}$.  The magnetic permeability (\ref{permeabilityQCD0})  in  the vacuum state (\ref{chomomagneticcondensate0}) is equal to  zero:
\beqa
\mu^{QCD}_{vac} &=& 1+ { b\ g^2  \over 96 \pi^2} \log{\langle 2 g^2 \CF \rangle_{vac}  \over \mu^4} ~ =0.
\eeqa
It is useful to derive the expression of the effective Lagrangian by using the renormalisation group equation \cite{Savvidy:1977as,Matinyan:1976mp}.
The solution of the renormalisation group equation in terms of effective coupling constant $\bar{g}(g,t)$, with the boundary condition $\bar{g}(g,0) =g$,  has the following form \cite{Savvidy:1977as,Matinyan:1976mp}:
\be\label{effectivecouplingel1}
{\partial \CL \over \partial \CF} = - {g^2 \over \bar{g}^2(t)},~~~~~~~~~{d \bar{g} \over dt } = \beta(\bar{g})~,~~~~~~~~t = {1\over 2}\ln(2 g^2 \CF/ \mu^4) .
\ee
 The derivative (\ref{effectivecouplingel1})  of the effective Lagrangian  has transparent expression in terms of the effective coupling constant  and allows to obtain  the effective Lagrangian by integration over $\CF$
in all order of the perturbative expansion: 
\be\label{fulleffeclagr1}
\CL(\CF)  = - \mu^4   \int  {e^{2 t} \over \bar{g}^2(t) } d t  ~,~~~~~~~~t = {1\over 2}\ln(2 g^2 \CF/ \mu^4), 
\ee
and find out the expressions  for the physical quantities beyond the  one-loop approximation. One can calculate different observables of physical interest that will include  the effective energy momentum tensor, vacuum energy density, the magnetic permeability, the  effective coupling constants  and their behaviour as a function of the external fields. In particular, the energy momentum tensor (\ref{tmunu}) will take the following form:
\beqa\label{allloopenergymomentum1}
T_{\mu\nu}   
&=&-\Big(G_{\mu\lambda} G_{\nu\lambda}-g_{\mu\nu} {1\over 4} 
G^2_{\lambda\rho} \Big) {g^2 \over \bar{g}^2(t)} + g_{\mu\nu}   \Big( \int  { e^{2 t} \over \bar{g}^2(t) } dt -  {1\over 2}{e^{2t} \over  \bar{g}^2(t)}     
\Big)  \mu^4.
\eeqa
The vacuum energy density can be expressed in terms of  the trace $T_{\mu\mu}$ as: 
\be\label{energytracerel1}
\epsilon =T_{00}  =  {\vec{\CH_a}^2 \over 2} {g^2 \over \bar{g}^2(t)} + {1\over 4} T_{\mu\mu} ,~~~~~~~\CG = 0,
\ee
where the trace of the energy momentum tensor $T_{\mu\mu}$ is given by the following expression:  
\beqa\label{traceel2}
T_{\mu\mu}  =  4 \mu^4  \int     { e^{2 t}   \beta(\bar{g}(t))  \over \bar{g}(t)^3} d t ~,~~~~~~~~t = {1\over 2}\ln(2 g^2 \CF/ \mu^4). 
\eeqa
The last formula  provides  all-loop expression
for the conformal anomaly in gauge field theories\footnote{If one considers the approximation in which $\bar{g}(t)$ is field independent $\bar{g}(t) \equiv g$ then (\ref{traceel2})   will reduce to the one given in literature \cite{Adler,Nielsen,Minkowski,Duff:1977ay}.}. As far as the beta function $\bar{\beta}(g)$ has no zeros,  is negative  analytical function of the coupling constant  and 
\be
\int^{\infty}_{g} {d g \over \beta(g)} < \infty,
\ee
the minimum of the energy density curve is defined by the extremum, where the derivative (\ref{effectivecouplingel1}) vanishes. It follows that the value of the chromomagnetic condensate   is \cite{Savvidy:1977as}
\be\label{chomomagneticcondensaterenormgroup}
\langle 2  g^2 \CF  \rangle_{vac}=    \mu^4  \exp{\Big(2 \int^{\infty}_{g(\mu)}{d g \over \beta(g)}\Big)}.
\ee
Considering the value of the field strength tensor $\CF_{0}$ at which the vacuum energy density (\ref{energyexpYM0}) vanishes  $\epsilon(\CF_0)=0$,  the point $\CF_{0}$ shown on  Fig.\ref{fig2},
one can observe  that the effective coupling constant (\ref{effectivecouplingel1}) at this field strength has the value ${96 \pi^2   \over   11 N }$  and tends to zero as $N \rightarrow \infty$.   The energy density curve $\epsilon(\CF)$ (\ref{energyexpYM0}) intersect the horizontal zero energy line  at the nonzero angle  $\theta >~ 0$
(see (\ref{intersectionangle}) and Fig.\ref{fig2}).   The energy density curve can be continuously extended from the point $\CF_0$  deep into the negative energy density region arbitrary close to the value of the vacuum condensate  $\langle \CF \rangle_{vac}$ by considering a  larger values of $N$ and keeping  the t'Hooft coupling constant $g^2 N$ small and fixed.  This  demonstrates that the true vacuum of the Yang-Mills theory is below the perturbative vacuum  and that there is a nonzero energy gap between perturbative and true vacuum states.     
  
The article is organised as follows. In the second section we shall use gauge and renormalisation group  invariant  scheme (\ref{renormcondition1})  \cite{PhDTheses,Savvidy:1977as} to renormalise the massless Heisenberg-Euler Lagrangian and derive the exact one-loop expression for the effective Lagrangian in QED (\ref{QED0}). In the third section we shall use the renormalisation group equations for the  effective Lagrangian to derive all loops results for the vacuum energy density and the traces of the energy momentum tensor. In the forth and fifth sections the analyses will be extended to the Yang-Mills theory and the phenomena of the chromomagnetic gluon condensation will be reexamined. We shall discuss the absence of the imaginary part in the YM effective Lagrangian in chromomagnetic field and the stability of the chromomagnetic gluon condensate by Niels Bohr theory group and by Kurt Flory.

\section{\it Heisenberg-Euler  Effective Lagrangian in Massless Limit}

The effective action $\Gamma$ and the effective Lagrangian $\CL$ in gauge field theories can be 
represented as a sum of the one-particle irreducible loop diagrams $W^{(n)}, n=1,2...$ :  
\be\label{allloop}
\Gamma = \int \CL~ d^4 x= S + W^{(1)} + W^{(2)}+....,
\ee
where $S$ is the Yang-Mills or  Maxwell action:
\be
S = - {1\over 4} \int G^2_{\mu\nu} d^4 x~.
\ee
We shall analyse the behaviour of the effective actions in both theories and  shall consider first the Quantum Electrodynamic in massless limit ( see, in particular, \cite{He:2014cra}). The Heisenberg-Euler Lagrangian in QED \cite{Sauter:1931zz,Heisenberg:1934pza,Euler:1935zz,Heisenberg:1935qt,Schwinger:1951nm,Coleman:1973jx,Dunne:2004nc} is a sum of the one-loop diagrams with a vacuum 
electron-positron pair running in the loop:
\be\label{lagrangian}
W^{(1)} = \int \CL^{(1)}(x) d^4 x~ ,
\ee 
and can be expressed through the 
functional determinant of the Dirac operator \cite{Schwinger:1951nm}:
\be
W^{(1)} = -i~ Tr~ ln (\gamma \Pi +m)= i \int^{\infty}_{0} {ds \over s} Tr e^{-i (\gamma \Pi +m) s}=
{i\over 2} \int^{\infty}_{0} {ds \over s} Tr e^{-i (m^2 -(\gamma \Pi)^2 ) s},
\ee
where $\Pi_{\mu} = -i \partial_{\mu} -e A_{\mu}$. The general expression for the one-loop effective Lagrangian $\CL^{(1)}$ has therefore the following form:
\be
\CL^{(1)}(x) = {i\over 2} \int^{\infty}_{0} {ds \over s}e^{-i m^2 s} tr (x \vert U(s) \vert x),
\ee
where 
\be
U(s) = e^{-i H s},~~~~ H = - (\gamma \Pi)^2 = \Pi^2_{\mu} - {1\over 2} e \sigma_{\mu\nu} F_{\mu\nu}.\nn
\ee
In the case of the constant electromagnetic field strength tensor $F_{\mu\nu }$ the matrix element 
of the operator $(x \vert U(s) \vert x)$ can be calculated exactly and has the following form \cite{Schwinger:1951nm}:
\be
(x \vert U(s) \vert x) = -{i\over (4\pi s)^2} e^{-L(s)} e^{{i\over 2} e \sigma F s}~,\nn
\ee
where 
\be
L(s) = {1\over 2} tr ln[(eFs)^{-1} \sinh(eFs)].\nn
\ee
The Lagrangian will take the form 
\be
\CL^{(1)} = {1\over 32 \pi^2} \int^{\infty}_{0} {ds \over s^3}e^{-i m^2 s} 
e^{-L(s)} tr e^{{i\over 2} e \sigma F s}. \nn
\ee
By the rotation of the integration contour in complex plane  $\textcircled{s}$ as $s \rightarrow -is$ one can get
\be\label{oneloopel}
\CL^{(1)} = -{1\over 32 \pi^2} \int^{\infty}_{0} {ds \over s^3}e^{- m^2 s} 
e^{-l(s)} tr e^{{1\over 2} e \sigma F s}~,
\ee
where
\be
l(s) = {1\over 2} tr ln[(eFs)^{-1} \sin(eFs)]. \nn
\ee
The traces in (\ref{oneloopel}) can be  evaluated by using the eigenvalues of the field strength 
tensor matrix $F_{\mu\nu} \psi_{\nu} = F \psi_{\mu}$. The characteristic equation is
\be
F^4  + 2 \CF F^2 - \CG^2=0,
\ee
where 
\be
\CF = {1\over 4} F_{\mu\nu}^2= {1\over 2}(\vec{\CH}^2 - \vec{\CE}^2),~~~
\CG = {1\over 4}  F_{\mu\nu} F^*_{\mu\nu} = \vec{\CE} \vec{\CH},
\ee
and has the solutions 
\be
F^2_{1}= -\CF - (\CF^2 +\CG^2)^{1/2},~~~~~F^2_{2}= -\CF + (\CF^2 +\CG^2)^{1/2}.\nn
\ee
Thus 
\be
e^{-l(s)} = { e F_{1} s ~e F_{2} s \over \sin (e F_{1} s) \sin (e F_{2}s )},~~~~
tr e^{{1\over 2} e \sigma F s} =  4 \cos (e F_{1} s) \cos (e F_{2}s).\nn
\ee
The Lagrangian (\ref{oneloopel}) will take the following form  \cite{Heisenberg:1935qt,Schwinger:1951nm}:
\be
\CL^{(1)} = -{1\over 8 \pi^2} \int^{\infty}_{0} {ds \over s^3}e^{- m^2 s} 
{ (e F_{1} s)  \cos (e F_{1} s) ~(e F_{2} s) \cos (e F_{2}s) \over \sin (e F_{1} s)~~~ \sin (e F_{2}s )} , \nn
\ee
and with real eigenvalues $f_1 =-i F_{1}, f_2 = F_2$ one can get
\be
\CL^{(1)} = -{1\over 8 \pi^2} \int^{\infty}_{0} {ds \over s^3}e^{- m^2 s} 
{ (e f_{1} s)  \cosh (e f_{1} s) ~(e f_{2} s) \cos (e f_{2}s) \over \sinh (e f_{1} s)~~~ \sin (e f_{2}s )}, 
\ee
where 
\be
f_1 = ( (\CF^2 +\CG^2)^{1/2} + \CF)^{1/2},~~~~f_2 = ((\CF^2 +\CG^2)^{1/2} -\CF )^{1/2}~.
\ee
In pure magnetic field configurations $\CG=0, \CF >0$ one can get  $f_1= (2 \CF)^{1/2}, f_2=0$ and  in pure electric case 
$\CG=0, \CF < 0$,  thus  $f_1=0, f_2=(-2 \CF)^{1/2}$:
\beqa\label{fields}
&&\CG=0,~ \CF >0~~~~~~f_1= (2 \CF)^{1/2}, f_2=0~~~~~~~~~  \text{pure magnetic}\\
&&\CG=0,~ \CF < 0~~~~~~ f_1=0, f_2=(-2 \CF)^{1/2} ~~~~~~~ \text{pure electric}.\nn
\eeqa
We shall consider the QED in the massless limit and impose the following renormalisation 
condition on the effective Lagrangian  introduced in \cite{Savvidy:1977as,Matinyan:1976mp}\footnote{This renormalisation scheme is alternative to the standard MS and other schemes, see, in particular,  \cite{Nielsen:1978zg}. }:
  \be\label{renormcondition}
 {\partial  \CL \over \partial \CF} \vert_{t = {1\over 2}\ln ({2e^2  \vert \CF \vert \over \mu^4})= \CG=0} =-1, 
 \ee
 where $\mu^2$ is the renormalisation scale parameter. This condition defines the renormalisation of the effective Lagrangian in a covariant gauge
$
\CL_{r} =  \CL_{un} - Z ~\CF.\nn
$
 In the case of pure magnetic field (\ref{fields}) the Lagrangian $ \CL^{(1)}$ has the following form:
 \be
 \CL^{(1)} = -{1\over 8 \pi^2} \int^{\infty}_{0} {ds \over s^3} 
{ (e f_1 s)  \cosh (e f_1 s)   \over \sinh (e f_1 s) } ,\nn
 \ee
and diverges at the boundaries of the proper time integration region. With the use of the renormalisation condition (\ref{renormcondition}) one can handle both divergences \cite{Savvidy:1977as,Matinyan:1976mp}. This leads to the following renormalisation of  the Heisenberg-Euler Lagrangian in the massless limit: 
 \beqa\label{masslessHE}
 \CL^{(1)} &=& -{ \mu^4 \over 8 \pi^2} \int^{\infty}_{0}  {ds \over s^3} 
\Big( {   a s  \cosh(a s) \over  ~ \sinh( a s) } -1- {a^2 s \over 2 }  \Big({\cosh s \over \sinh s } - {s  \over \sinh^2 s}\Big) \Big),
 \eeqa
where 
\be
a^2 = 2 e^2 \CF /\mu^4  = e^2 \vec{\CH}^2 /\mu^4  
~~~~~~~, ~~\CG=0.
\ee
One can get convinced that this expression is well defined in both limits, in the ultraviolet $s \rightarrow 0$ and in the infrared $s \rightarrow \infty$ regions. One can calculate this integral exactly. 
The integrals appearing in this expression can be expressed in terms of the Riemann zeta function and its extension (see the Appendix for details).   
The Lagrangian (\ref{masslessHE}) will take the following form:
\beqa
 \CL^{(1)}_k &=& -{ \mu^4 \over 8 \pi^2} \int^{\infty}_{0}  ds   
\Big( {   a s^{k-1}  \cosh(a s) \over  ~ \sinh( a s) } -   {a^2 s^{k-1} \cosh s \over 2 \sinh s } 
+  { a^2 s^{k}  \over 2 \sinh^2 s} \Big)=\nn\\
&=& -{ \mu^4 \over 8 \pi^2} 
\Big[2 a^{1-k}  -   a^2  
+    a^2  k  \Big] {\Gamma(k)  \zeta(k) \over 2^k}, \nn
 \eeqa
and in the limit $k \rightarrow -1$ we shall get 
\beqa
 \CL^{(1)} 
&=& { \mu^4   a^2 \over 4 \pi^2} 
\Big[  \ln a  - {1\over 2}   \Big]~  \lim_{k \rightarrow -1} ~  {(k+1) \Gamma(k)  \zeta(k) \over 2^k}=
 { \mu^4   a^2 \over 24 \pi^2} 
\Big[  \ln a  - {1\over 2}   \Big],\nn
\eeqa
where we used the identity \cite{gradshteyn}
$
\lim_{k \rightarrow -1} ~ (k+1) \Gamma(k)  \zeta(k) = \lim_{k \rightarrow -1} {2^{k-1} \pi^k \zeta(1-k) \over  \cos({\pi k \over 2})}= 
{1\over 6}. \nn
$
Thus, in terms of Lorentz and gauge invariant  $\CF= {1\over 4} F_{\mu\nu}^2 $, the exact expression of the one-loop 
Lagrangian in massless  QED is:
\beqa
 \CL^{(1)}_{el} =  {   e^2 \CF  \over 24 \pi^2} 
\Big[  \ln ({2 e^2 \CF \over \mu^4})  - 1  \Big] ,~~~~~~~~
\eeqa
where  $2 \CF = \vec{\CH}^2 - \vec{\CE}^2 > 0, ~ 
\CG = \vec{\CE} \vec{\CH} =0$  and the effective Lagrangian will take the following form:
 \beqa\label{effectivelagrangianel}
 \CL &=& -\CF + {   e^2 \CF  \over 24 \pi^2} 
\Big[  \ln ({2 e^2 \CF \over \mu^4})  - 1  \Big].
\eeqa
As it follows from this expression, the  QED vacuum responds to the background magnetic field by inducing a vacuum  current of the electron-positron pairs, which attenuates the magnetic field imposed on the vacuum. 
The magnetic induction $\vec{\CB}$  of the QED vacuum  is \cite{PhDTheses}: 
\be\label{palar}
\vec{\CB} = - {\partial \CL \over \partial \vec{\CH}} 
= \vec{\CH} \Big[1- {e^2  \over 24 \pi^2}\log{e^2 \vec{\CH}^2 \over \mu^4} \Big]~~~= \mu_{vac}~   \vec{\CH},
\ee
and the QED vacuum responds to the background magnetic field as a diamagnet  with the magnetic permeability of the following form:
\be\label{permeability}
\mu_{vac} = 1- {e^2  \over 24 \pi^2}\log({e^2 \vec{\CH}^2 \over \mu^4}) ~< 1~~~~~\text{\it diamagnetic}.
\ee
The diamagnetism of the QED vacuum means that it repels the magnetic fields by forming induced magnetic field in the direction opposite to that of the applied magnetic field. This phenomenon is similar to  the Landau diamagnetism of free electron gas when the counteracting field is formed when the electron trajectories are curved due to the Lorentz force.  This also can be seen from the vacuum energy expression  ( see Fig.\ref{fig1} ):
\be\label{energyexp}
\epsilon= {\vec{\CH}^2 \over 2} - {e^2 \vec{\CH}^2 \over 48 \pi^2}
[\log({e^2 \vec{\CH}^2 \over \mu^4})- 1].
\ee
In the case of pure electric field the one-loop Lagrangian has the following form:
\beqa\label{masslessHE1}
 \CL^{(1)} &=& -{ \mu^4 \over 8 \pi^2} \int^{\infty}_{0}  {ds \over s^3} 
\Big( {   b s  \cos(b s) \over  ~ \sin( b s) }-1 - {b^2 s \over 2 }  \Big({\cos s \over \sin s } - {s  \over \sin^2 s}\Big) \Big),
 \eeqa
where 
\be
b^2 = -2 e^2 \CF  /\mu^4  = e^2 \vec{\CE}^{\ 2} /\mu^4  
~~~~~~~, ~~\CG=0,
\ee
and has  singularities at $s = s_n = \pi n /b = \mu^2 \pi n /e \CE$. The integration path is considered to lie above the real axis, therefore  we shall obtain a large positive imaginary contribution to $\CL^{(1)}$\footnote{ The universal character of the electric instability of the vacuum was discussed  in the recent article \cite{Pimentel:2018fuy}.}:
\be\label{qedimaginary}
2 Im \CL^{(1)} = {e^2 \CE^2 \over 4 \pi^3} \sum^{\infty}_{n=1} {1 \over n^2}=  {e^2 \CE^2 \over 24 \pi} ~. 
\ee
The real part of the Lagrangian in the electric field is 
\be\label{lagrangianexpelectric}
Re  \CL^{(1)} =-{e^2 \vec{\CE}^2 \over 48 \pi^2}
[\log({e^2 \vec{\CE}^2 \over \mu^4})- 1].
\ee
The formulas (\ref{effectivelagrangianel}), (\ref{lagrangianexpelectric}) and (\ref{qedimaginary}) prove that the effective Lagrangian is the analytical function of the variable $\CF$ and has the 
general form (\ref{effectivelagrangianel}).
The corresponding energy density takes the following form: 
\be
 \epsilon= {\vec{\CE}^2 \over 2} - {e^2 \vec{\CE}^2 \over 48 \pi^2}
[\log({e^2 \vec{\CE}^2 \over \mu^4})+ 1],
\ee
and its behaviour is similar to the one shown on Fig.\ref{fig1}. The electric permeability $\vec{\CD} =  {\partial \CL \over  \partial \vec{\CE}} = \varepsilon   \vec{\CE}$, where 
\be
\varepsilon = 1 -  {e^2  \over 24 \pi^2} \log({e^2 \vec{\CE}^2 \over \mu^4}) +i {e^2  \over 24 \pi}.
\ee
In the next section we shall consider the renormalisation group invariant derivation of the all-loop effective Lagrangian (\ref{allloop}) and the generalised expressions for the magnetic induction (\ref{palar}) and permeability (\ref{permeability}) as well as the electromagnetic energy-momentum tensor and its trace.
 
\section{\it Renormalisation Group Equation for  Effective Lagrangian }
Let us derive the exact expression of the effective Lagrangian by using the renormalisation group equation \cite{Savvidy:1977as,Matinyan:1976mp}. The effective action $\Gamma$ is renormalisation group invariant quantity:  
\beqa
 \Gamma &=& \sum_n \int dx_1...dx_n \Gamma^{(n) a_1...a_n}_{~~~\mu_1...\mu_n}(x_1,...,x_n) A^{a_1}_{\mu_1}(x_1)...
 A^{a_n}_{\mu_n}(x_n),  \nn
 \eeqa 
because the vertex functions and gauge fields  transforms as follows: 
\beqa
\Gamma^{(n)~ a_1...a_n}_{r~~~\mu_1...\mu_n}  = Z^{n/2}_{3}\Gamma^{(n)~~a_1...a_n}_{un~~~\mu_1...\mu_n},~~~~~
A^{a}_{\mu}(x)_r = Z^{-1}_{3}A^{a}_{\mu}(x)_{un},~~~~
g_r = Z^{1/2}_{3} g_{un} .~~~\nn
\eeqa
The renormalisation group equation takes the form 
\be
\{ \mu^2 {\partial \over \partial \mu^2} + \beta(g) {\partial \over \partial g} + \gamma(g) 
\int d^4x A^{a}_{\mu}(x) {\delta \over \delta A^{a}_{\mu}(x)} \} \Gamma =0, \nn
\ee
where $\beta(g)$ is the Callan-Symanzik beta function, the $\gamma(g)$ is the anomalous dimension.   When  $\CG = \vec{\CE} \vec{\CH} =0 $ it reduces to the form
\be
\{ \mu^2 {\partial \over \partial \mu^2} + \beta(g) {\partial \over \partial g} + 2 \gamma(g) 
\CF {\partial \over \partial \CF}  \}  \CL =0,\nn
\ee
where in the covariant background  gauge $\beta = - g \gamma$ \cite{PhDTheses}. By introducing a dimensionless quantity  
\be\label{derivative}
\CM(g,t) = {\partial \CL \over \partial \CF},~~~~~~~~~t = {1\over 2}\ln(2 g^2 \CF/ \mu^4)  
\ee
one can get 
\be\label{renormequation}
\{ - {\partial \over \partial t} + \bar{\beta}(g) {\partial \over \partial g} + 2 \bar{\gamma}(g)  \}  \CM(g,t) =0, 
\ee
where 
\be\label{fullbeta}
\bar{\gamma} =  {\gamma \over 1- \gamma} ,~~~~~~~\bar{\beta} =  {\beta \over 1- \gamma}
\ee
and (\ref{renormcondition}) plays the role of the boundary condition:
\be\label{boundarycondition}
\CM(g,0) = -1.
\ee
From equations (\ref{renormequation}) and (\ref{boundarycondition}) it follows that 
\be\label{beta}
\bar{\gamma}=  -{1\over 2}{\partial \CM(g,t) \over \partial t} \vert_{t=0}, ~~~~
\bar{\beta}=   {1\over 2} g {\partial \CM(g,t) \over \partial t} \vert_{t=0}~.
\ee
The solution of the renormalisation group equation (\ref{renormequation}) in terms of effective coupling constant $\bar{g}(g,t)$, with the boundary condition $\bar{g}(g,0) =g$,  has the following form \cite{Savvidy:1977as,Matinyan:1976mp}:
\be\label{effectivecouplingel}
  {\partial \CL \over \partial \CF} = - {g^2 \over \bar{g}^2(t)},~~~~~~~~~{d \bar{g} \over dt } = \bar{\beta}(\bar{g}) .
\ee
The behaviour of the effective Lagrangian at large fields is similar to the behaviour of the gauge theory at large momentum. 
It follows that $\CM(g,t)$ is completely determined for all $t$ in terms of its first derivative (\ref{beta}) at $t=0$. To define the effective Lagrangian $\CL$ one should perform additional integration, which we shall do in the next section.

The above results allow to obtain renormalisation group expressions  for the physical quantities 
considered above in one-loop approximation. Indeed, with these expressions in hand we can calculate different observables of physical interest, that will include  the effective energy momentum tensor, vacuum energy density, the magnetic permeability, the  effective coupling constants  and their behaviour as functions of the external fields. 

\section{\it Massless QED}

By using the one loop expression (\ref{effectivelagrangianel}) derived above
one can calculate the derivative 
\be\label{magnetisation}
 \CM(t,e) = {\partial \CL \over  \partial \CF } = -1 + {   e^2   \over 24 \pi^2} 
  \ln {2 e^2 \CF \over \mu^4},~~~~~~~~~\CG=0,   
\ee
and the Callan-Symanzik beta function  (\ref{beta}) takes the following form:
\be
\bar{\beta}=  {1\over 2}e {\partial \CM \over \partial t} \vert_{t =0}= {1  \over 24 \pi^2} e^3~.
\ee
The effective coupling constant (\ref{effectivecouplingel}) in the one-loop approximation is
\be\label{effectivee}
\bar{e}^2(\vec{\CH}^2)= {e^2  \over 1 - {e^2 \over 2 4 \pi^2} \log({e^2 \vec{\CH}^2 \over \mu^4})}
\ee
and tends to infinity at the magnetic field 
\be 
e^2  \CH_0^2 =\mu^4 e^{24 \pi^2 \over  e^2}.
\ee
In order to estimate the value of the critical field one can consider the  mass parameter $\mu $  to be of the order of the electron mass $m$. Then one can get 
$$
H^2_0  = ({m^2 c^3 \over e \hbar})^2 \exp{({24 \pi^2 \over e^2/\hbar c})} =H^2_c \exp{({24 \pi^2 \over e^2/\hbar c})}, 
~~~~~\alpha = {e^2 \over 4 \pi  \hbar c}  ,
$$ 
where  the critical field $H_c$ is
$$
H_c = {m^2 c^3 \over e \hbar} \approx 4.4 ~ 10^{13} ~Gauss.
 $$
The perturbation expansion breaks down at the "Moscow zero"  shown on Fig.\ref{fig1}.

As far as the derivative (\ref{effectivecouplingel})  of the effective Lagrangian (\ref{derivative}) has transparent expression in terms of the effective coupling constant  (\ref{effectivecouplingel}) 
one  can obtain  the effective Lagrangian by integration over $\CF$:
\be
\CL(\CF)  =\int {\partial \CL \over \partial \CF}  d \CF = - \int {g^2  \over \bar{g}^2(t) }d \CF. 
\ee
By using the relation (\ref{derivative}) to express the differential $ g^2 d \CF = \mu^4  e^{2\tau} d\tau $ through $d\tau$  one can represent the Lagrangian in the form:
\be\label{fulleffeclagr}
\CL(\CF)  = - \mu^4   \int   {e^{2\tau} \over \bar{g}^2(\tau) } d\tau  ~,~~~~~~~~t = {1\over 2}\ln(2 g^2 \CF/ \mu^4) .
\ee 
In massless QED the magnetic induction  (\ref{palar}) will take the following form:
\beqa
\vec{\CB} = -  {\partial \CL \over \partial \vec{\CH}  } = -  {\partial \CL \over \partial \CF  }  ~  \vec{\CH}= 
{\it  \mu }_{vac}~  \vec{\CH}. 
\eeqa
Therefore the vacuum permeability (\ref{permeability}) can be expressed  through the effective coupling constant $\bar{e}^2(t)$:
\be\label{permeabilityfull}
{\it \mu}_{vac}~ = ~ {e^2 \over \bar{e}^2(t)}. 
\ee
The effective Lagrangian approach allows to calculate the quantum-mechanical corrections to the energy momentum tensor by using the formula derived by Schwinger in \cite{Schwinger:1951nm}:
\beqa\label{energymomentum}
T_{\mu\nu} &=& - g_{\mu\nu} \CL  + {\partial \CL  \over F_{\mu\lambda}} F_{\nu\lambda}
= - g_{\mu\nu} \CL + F_{\mu\lambda} F_{\nu\lambda} ~{\partial \CL \over \partial \CF} 
+ ~ g_{\mu\nu} {\partial \CL \over \partial \CG}\CG \nn\\
&=&(F_{\mu\lambda} F_{\nu\lambda}-g_{\mu\nu} {1\over 4} 
F^2_{\lambda\rho} ) {\partial \CL \over \partial \CF}- g_{\mu\nu} (\CL -
\CF {\partial \CL \over \partial \CF} - \CG {\partial \CL \over \partial \CG}).
\eeqa
In our case, when $\CG = 0$, we shall find all-loop expression for $T_{\mu\nu}$ by using (\ref{fulleffeclagr}):
\beqa\label{allloopenergymomentum}
T_{\mu\nu}   
&=&-\Big(F_{\mu\lambda} F_{\nu\lambda}-g_{\mu\nu} {1\over 4} 
F^2_{\lambda\rho} \Big) {e^2 \over \bar{e}^2(t)} + g_{\mu\nu}   \Big( \int   { e^{2 t} \over \bar{e}^2(t) } d t -  {1\over 2}{e^{2t} \over  \bar{e}^2(t)}     
\Big)  \mu^4 .
\eeqa
And for the vacuum energy density we shall get:  
\be\label{energytracerel}
\epsilon= T_{00} = {\vec{\CH}^2 \over 2} {e^2 \over \bar{e}^2(t)} + {1\over 4} T_{\mu\mu} = \mu^4   \int  {e^{2 t} \over \bar{e}^2(t) } d t ,
\ee
where the trace of the energy momentum tensor $T_{\mu\mu}$ is not equal to zero and characterises the breaking of conformal symmetry in massless QED:
\beqa\label{traceel}
T  &\equiv&  T_{\mu\mu} = 4(
\CF {\partial \CL \over \partial \CF}  - \CL ) = 4  \mu^4 \Big(   \int  {  e^{2 t} \over \bar{e}^2(t) } d t 
-  {1\over 2} { e^{2t} \over  \bar{e}^2(t)}  \Big) ~, 
\eeqa
where $\CG=0,~~t = {1\over 2}\ln(2 e^2 \CF/ \mu^4) .$
It is also useful to obtain the derivative of $T$ expressed in terms of the effective coupling constant: 
\beqa\label{traceel1}
{\partial T \over \partial \CF}    &=&  4   \CF  {\partial \CL  \over  \partial \CF^2}  = 
4   \CF  {\partial \CM  \over  \partial \CF} =  2    {\partial \CM  \over   \partial t} =
-2    {\partial  \over   \partial t}~  {e^2 \over \bar{e}^2(t)}, \nn
\eeqa
and by using (\ref{effectivecouplingel}) we shall get
\be\label{renormtraceenergy}
{\partial T \over \partial \CF} =  4 e^2~ {\bar{\beta}(\bar{e})  \over \bar{e}^3}~.
\ee
The integration of (\ref{renormtraceenergy})  over $\CF$  provides the alternative  forms of  (\ref{traceel}): 
\be 
T = 4 \mu^4  \int^{t}_{-\infty}     { e^{2 t}  \bar{\beta}(\bar{e}(t))  \over \bar{e}^3(t)} d t =4 \mu^4  \int      { e^{2 t}  d  \bar{e}(t)   \over \bar{e}^3(t)}   = -2 \mu^4  \int    e^{2 t} d {1\over \bar{e}^2(t)} .
\ee
The last two formulas (\ref{traceel}) and (\ref{traceel1}) provide the all-loop expressions 
for the conformal anomaly in QED in the massless limit. If one considers the approximation in which $\bar{e}(t)$ is field independent $\bar{e}(t) \equiv e$ then (\ref{traceel}), (\ref{traceel1})  will reduce to the expression $T = 2 {\beta(g) \over g} \CF$ given in literature \cite{Adler,Nielsen,Minkowski}.

For the one-loop energy momentum tensor (\ref{energymomentum})
  we shall get  
\beqa
T_{\mu\nu} = T^{el}_{\mu\nu}\Big[1 - {   e^2   \over 24 \pi^2} 
  \ln {2 e^2 \CF \over \mu^4}   \Big]
+ g_{\mu\nu}    {   e^2   \over 24 \pi^2} \CF  ,~~~~~~~~~\CG=0, 
\eeqa
where we used the expressions (\ref{magnetisation}). The energy density and the trace of the energy momentum tensor are  
\beqa\label{vacuumenergy}
T_{00}=  {\vec{\CH}^2 \over 2}\Big(1 - {e^2 \vec{\CH}^2 \over 24 \pi^2}
\log{e^2 \vec{\CH}^2 \over \mu^4} \Big) +  {e^2   \over 24 \pi^2}{ \vec{\CH}^2 \over 2}, ~~~~~~~~~~~~T_{\mu\mu}  =     {   e^2   \over 6 \pi^2} { \vec{\CH}^2 \over 2} 
\eeqa 
and they represent the one-loop approximation  of  (\ref{energytracerel}) and  (\ref{traceel}). In the next section we shall consider the behaviour of the effective  Lagrangian in Yang-Mills theory and QCD. 
 
\section{\it Effective Lagrangian of  Yang-Mills theory}
The loop expansion  of the effective action in Yang-Mills theory has the following form:
 \beqa\label{loopexpansion}
 \Gamma &=& \sum_n \int dx_1...dx_n \Gamma^{(n) a_1...a_n}_{~~~\mu_1...\mu_n}(x_1,...,x_n) A^{a_1}_{\mu_1}(x_1)...
 A^{a_n}_{\mu_n}(x_n) \nn\\
 &=& S_{YM} + W^{(1)} + W^{(2)} +....,
 \eeqa 
and the one-loop effective Lagrangian has the form \cite{Vanyashin:1965ple,Skalozub:1975ab,Brown:1975bc,Duff:1975ue,PhDTheses,Batalin:1976uv}
\be\label{oneloopeffec}
 W^{(1)} =  S_{YM}(A) + {i \over 2}  Tr \ln [{\delta^2 S_{YM}(A) \over \delta A~ \delta A}] -
i Tr \ln [\nabla_{\mu}(A) \nabla_{\mu}(A)],
\ee
where 
\beqa
S_{YM}(A) &=& -  {1 \over 4} \int d^4 x~ tr G_{\mu\nu}G_{\mu\nu},~~~~ ~~~G_{\mu\nu}= \partial_{\mu} A_{\nu}-\partial_{\nu} A_{\mu}  - i g [ A_{\mu},  A_{\mu}]
\nn\\ \label{operator}
H_{\mu\nu}(\alpha)= {\delta^2 S_{YM}(A) \over \delta A ~\delta A} &=& \eta_{\mu\nu} \nabla_{\sigma}(A) \nabla_{\sigma}(A) -2 g G_{\mu\nu} +
(\alpha -1)\nabla_{\mu}(A) \nabla_{\nu}(A),\\
 H_{FP}&=& \nabla_{\mu}(A) \nabla_{\mu}(A).\nn
\eeqa
By using proper time representation we shall get the effective action in the following form:
\be
\Gamma(A) =  S_{YM}(A) -  {i \over 2}  \int^{\infty}_{0} {ds \over s }Tr e^{-i H(\alpha) s }   +
i  \int^{\infty}_{0} {ds \over s }Tr  e^{-i H_{FP} s }, 
\ee
and for the effective Lagrangian the following expression: 
\be
\CL_{eff} =  \CL_{YM} -  {i \over 2}  \int^{\infty}_{0} {ds \over s }Tr (x\vert U(s) \vert x)   +
i  \int^{\infty}_{0} {ds \over s }Tr   (x\vert U_{0}(s) \vert x), 
\ee
where 
$
 U(s)   = e^{-i H(\alpha) s } ,~U_{0}(s) = e^{-i H_{FP} s }~.
$
The Green function in the background field has the following form: 
\be
G(x,y;A) =  -i  \int^{\infty}_{0} ds ~ (x\vert U(s) \vert y) . 
\ee
As is was proven in \cite{PhDTheses,Batalin:1976uv}, the  $\CL_{eff} $ is $\alpha$ independent functional on the solutions of the  YM classical equations. On the covariantly constant gauge field solution \cite{Brown:1975bc,Duff:1975ue,Batalin:1976uv,PhDTheses}
\be
A^a_{\mu}= -{1\over 2} G_{\mu\nu} x_{\nu} n^a,~~~n^2=1, ~~~~x_{\mu} A^a_{\mu}=0
\ee
the matrix elements can be calculated and have the following form  \cite{Batalin:1976uv,PhDTheses}:
\be
(x\vert U(s) \vert y) = {i \over (4\pi s)^2} \exp{\{-{i\over 4} (x-y) K(s) (x-y) + {i\over 2} x N y - L(s) + 2 N s}\}
\ee
\be
(x\vert U_0(s) \vert y) = {i \over (4\pi s)^2} \exp{\{-{i\over 4} (x-y) K(s) (x-y) + {i\over 2} x N y  - L(s)}\},
\ee
where the corresponding matrices are
\beqa
N &=& i g G\nn\\
K(s) &=& N \coth (N s) \nn\\
L(s) &=& {1\over 2} tr \ln [(N s) \sinh(N s) ]
\eeqa
and 
\be
\CL^{(1)}  = - {1\over 32 \pi^2} \int   {ds \over s^3} Tr \exp{\{ - L(s) + 2 N s\}}
+ {1\over 16 \pi^2} \int   {ds \over s^3} Tr \exp{\{ - L(s) \}}.
\ee
By substituting the matrix elements and calculating the traces one can get \cite{PhDTheses,Batalin:1976uv}:
\beqa
\CL^{(1)} = &-&{1\over 8 \pi^2} \int   {ds \over s^3} e^{-i \mu^2 s}
{ (g F_{1} s)   ~(g F_{2} s)   \over \sinh (g F_{1} s)~ \sinh (g F_{2}s )} -\nn\\
&-&{1\over 4 \pi^2} \int   {ds \over s^3} e^{-i \mu^2 s} (g F_{1} s) ~(g F_{2} s)
\Big[{   \sinh (g F_{1} s)  \over \sinh (g F_{2} s) } + {   \sinh (g F_{2} s)  \over \sinh (g F_{1} s) }\Big], 
\eeqa
where 
\be
F^2_{1}= -\CF - (\CF^2 +\CG^2)^{1/2},~~~~~F^2_{2}= -\CF + (\CF^2 +\CG^2)^{1/2}~.
\ee
The first integral  here coincides, up to the coefficient 2, with the expression of the one-loop  Lagrangian in the scalar electrodynamics. The doubling of this expression is associated with the additional degrees of freedom due to the vector bosons isospin. The second term is due to the spin contribution  $- 2 g G_{\mu\nu}$  in the operator $H_{\mu\nu}$. We introduced the mass parameter $\mu^2$ in order to control  the infrared singularities and  to make the integrals convergent at infinity \cite{PhDTheses}. 
Still, this is not enough to make integrals convergent at infinity. By using the real eigenvalues 
\be
f^2_1 = \CF + (\CF^2 +\CG^2)^{1/2}, ~~~~~~~~~f^2_{2}= -\CF + (\CF^2 +\CG^2)^{1/2}
\ee
one can observe that the second term in  the square bracket will take the form ${   \sinh (g f_{2} s)  \over \sin (g f_{1} s) } $ and the integral diverges exponentially in the infrared region at infinity. We shall choose the integration counter in the complex  plane $\textcircled{s} $ so as to guarantee the convergence of the last integral. For that one should rotate  the integration counter in the  third integral by the substitution $s \rightarrow  - i s$. The same rotation of the counter can be performed in the first integral as far it is convergent in any way. Thus we shall get  \cite{Batalin:1976uv,PhDTheses}
\beqa
\CL^{(1)} = &&{1\over 8 \pi^2} \int^{\infty}_{0}    {ds \over s^3} e^{- \mu^2 s}
{ (g f_{1} s)   ~(g f_{2} s)   \over \sinh (g f_{1} s)~ \sin (g f_{2} s )} +\nn\\
&+&{1\over 4 \pi^2} \int^{\infty}_{0}  {ds \over s} e^{-i \mu^2 s} (g f_{1} ) ~(g f_{2} )
{   \sin (g f_{1} s)  \over \sinh (g f_{2} s) } \nn\\
&-& {1\over 4 \pi^2} \int^{\infty}_{0}   {ds \over s} e^{- \mu^2 s} (g f_{1} ) ~(g f_{2} )  {   \sin (g f_{2} s)  \over \sinh (g f_{1} s) }. 
\eeqa
The integrals are still diverging in the ultraviolet region at the $s=0$.
In order to renormalise the Lagrangian we have to identify the ultraviolet divergences in the above integrals. These are
\beqa
&& { (g f_{1} s)   ~(g f_{2} s)   \over \sinh (g f_{1} s)~ \sin (g f_{2} s )}     = 1 -{g^2\over 6} (f^2_1 -f^2_2) s^2  +\CO(s^4) \nn\\
&& f_{1}f_{2}{  \sin (g f_{1} s )   \over \sinh (g f_{2} s)~}     = f^2_{1}   +\CO(s^2) \nn\\
&& f_{1}f_{2}{  \sin (g f_{2} s )   \over \sinh (g f_{1} s)~}     = f^2_{2}   +\CO(s^2). \nn
 \eeqa
Subtracting these terms, which are quadratic in the field strength tensor, we shall get the renormalised effective Lagrangian  \cite{PhDTheses}:
\beqa\label{effYMlagrangian}
\CL^{(1)} = &&{1\over 8 \pi^2} \int^{\infty}_{0}  {ds \over s^3} e^{- \mu^2 s}
\Big( { (g f_{1} s)   ~(g f_{2} s)   \over \sinh (g f_{1} s)~ \sin (g f_{2} s )} -1 + {1\over 6}( g s)^2 (f^2_1 -f^2_2) \Big)+\nn\\
&+&{g^2 \over 4 \pi^2} \int^{\infty}_{0} {ds \over s} e^{-i \mu^2 s}   \Big(  f_{1} f_{2} 
{   \sin (g f_{1} s)  \over \sinh (g f_{2} s) }  - f^2_1 \Big) \nn\\
&-& {g^2 \over 4 \pi^2} \int^{\infty}_{0} {ds \over s } e^{- \mu^2 s}  \Big( f_{1} f_{2} {   \sin (g f_{2} s)  \over \sinh (g f_{1} s) }  - f^2_2\Big).
\eeqa
Now the integrals are convergent in both regions, in the infrared and in the ultraviolet. 
First let us consider a pure chromomagnetic  field:  
$$\CG=0,~~ \CF = {\vec{\CH_{a}}^2 - \vec{\CE_{a}}^2  \over 2} > 0,~~~~ 
f^2_1 = 2 \CF ,~~ f^2_2 =0.$$
The Lagrangian (\ref{effYMlagrangian}) will take the form
\beqa\label{effmagneticlag}
\CL^{(1)} = &+&{1\over 8 \pi^2} \int^{\infty}_{0}  {ds \over s^3} e^{- \mu^2 s}
\Big( {  g f_1 s    \over \sinh (g f_1 s) } -1 + {1\over 6}( g s)^2 f^2_1 \Big)+\nn\\
&+&{g^2 \over 4 \pi^2} \int^{\infty}_{0} {ds \over s} e^{-i \mu^2 s}   \Big( 
{ f_1  \sin (g f_1 s)  \over  g  s  }  - f^2_1).
\eeqa
 The asymptotic behaviour of the real part in chromomagnetic fields is \cite{PhDTheses} (page 49, (2.3.15)):
\beqa\label{asymptoticfree}
\Re \CL^{(1)} ~\approx ~+ {g^2  \over 48 \pi^2} {\vec{\CH}^2_a \over 2}\ln {g^2 \vec{\CH}^2_a \over \mu^4}~ - ~
{g^2  \over 4 \pi^2} {\vec{\CH}^2_a \over 2}\ln {g^2 \vec{\CH}^2_a \over \mu^4} ~= 
 -{11 g^2 \over 48 \pi^2}   {\vec{\CH}^2_a \over 2}\ln {g^2 \vec{\CH}^2_a \over \mu^4}, ~~~
\eeqa
where the first term represents the diamagnetism, which counteracts to the external field caused by the 
quantum current induced by the charged vector bosons in the vacuum and the second term represents the 
paramagnetism, an effect associated with the polarisation of the gluon spins, which, as one can see, dominates the asymptotic behaviour \cite{PhDTheses}. The imaginary part of the effective Lagrangian (\ref{effmagneticlag}) in background chromomagnetic field in our regularisation scheme  vanishes \cite{PhDTheses}:
\beqa\label{imaginarypart}
\Im \CL^{(1)} 
 &=& -  {g f_1 \over 4 \pi^2} \int^{\infty}_{0} {ds \over s^2}      
  \sin( \mu^2 s)  \sin (g f_{1} s)     + {g^2 f^2_1 \over 4 \pi^2} \int^{\infty}_{0} {ds \over s} \sin( \mu^2 s)    \nn\\
&&\nn  \\
  &=& -  {g f_1 \over 4 \pi^2}   {\pi \over 2 } g f_{1}      + {g^2 f^2_1 \over 4 \pi^2}  {\pi \over 2 }    
= -  {g^2 f^2_1  \over 8 \pi}      + {g^2 f^2_1 \over 8 \pi}   =0,    
\eeqa
where  $ f_1 = \sqrt{  \vec{\CH_a}^2   }= \CH$.  A similar conclusion was derived in \cite{Dittrich:1983ej} by using alternative regularisation. The significance of the absence/presence of the imaginary part in the effective Lagrangian connected with the fact that it defines the quantum-mechanical stability of a given field configuration.  The above conclusion on the absence of the  imaginary part is not conclusive  due to the negative-higgs-like eigenmode in the operator (\ref{operator})  \cite{Nielsen:1978rm,Skalozub:1978fy,Ambjorn:1978ff}, and we shall discuss the stability of the background field configurations in the subsequent eighth section. Here we shall refer to the articles of Ambjorn, Nielsen and Olesen \cite{Ambjorn:1978ff}, Leutwyler  \cite{Leutwyler:1980ev,Leutwyler:1980ma} and Flory \cite{Flory:1983dx,parthasarathy}, where they come to the same conclusion that there is no imaginary part in the effective Lagrangian in chromomagnetic  field.  Leutwyler  was considering the self-dual chromomagnetic   background field configurations and demonstrated that there is no imaginary part in the effective Lagrangian \cite{Leutwyler:1980ev,Leutwyler:1980ma} and that the effective Lagrangian has the form identical  to (\ref{YMeffective}).  Ambjorn, Nielsen and Olesen and Flory included  the quartic self-interaction of the higgs-like eigenmode which in physical terms corresponds to the summation of the contributions  coming from all loop diagrams  with higgs-like mode propagation in the loops.  The  physical reason behind this universality lies in the fact that even when the background field depends on space-time coordinates the part of the effective Lagrangian $\bar{\CL}(\CF,\CG)$  which depends only on the field strength tensor but not of its covariant derivatives has a  universal form \cite{PhDTheses,Savvidy:1977as,Matinyan:1976mp}.  In other words, as far as the wavelength of the fluctuating fields is very long, the effective action is not sensitive to the fine structure of the fluctuating fields.

Let us now consider pure chromoelectric  fields  $\CG=0, \CF   < 0$ and 
$f^2_1 = 0 , f^2_2 = - 2 \CF$:
\beqa
\CL^{(1)} = &&{1\over 8 \pi^2} \int^{\infty}_{0}  {ds \over s^3}  
\Big( {   ~(g f_{2} s)   \over  ~ \sin (g f_{2} s )} -1 + {1\over 3}( g s)^2 \CF \Big)+\nn\\
 &-& {g^2 \over 4 \pi^2} \int^{\infty}_{0} {ds \over s }  
 \Big(  f_{2} {   \sin (g f_{2} s)  \over g s }  - f^2_2\Big).\nn
\eeqa
The Lagrangian has singularities on the real axis at  $s_n = n \pi / e \CE$, and the integration path is considered to  lie above the real axis \cite{Batalin:1976uv,PhDTheses}:
\be\label{imagelectric}
2  Im \CL^{(1)} =   
{( g \CE)^2 \over 4 \pi^3 } \sum^{\infty}_{n=1}{(-1)^{n+1} \over n^2} =  { g^2 \CE^2 \over 48 \pi }
\ee
This is the probability per unit time and per unit volume that gluons are created by the chromoelectric  field. Due to the masslessness  of QCD  gluons the above formula does not contain the  Sauter-Schwinger exponentially small tunnelling factor $\exp{(-\pi {m^2 c^3 \over e \hbar \CE})}$ and even a weak  chromoelectric field will break down creating a cloud of soft gluons from the vacuum neutralising the imposed colour electric field.  While the exact results for the imaginary part of the effective action (\ref{imagelectric}) depend on the details of the background field, it was argued in \cite{Pimentel:2018fuy,Gies} that the threshold singularity is universal. The physical reason for this universality lies in the fact that the onset of pair production is dominated by the long-range fluctuations of the particles created   from the vacuum and becomes insensitive to the details of the field profile \cite{Pimentel:2018fuy,Gies}. 

The infrared long wavelength gluons  created from the vacuum are strongly interacting.  In the one-loop approximation the interaction between the produced gluons is not considered.  In \cite{Gyulassy:1986jq,Gyulassy:1986da} the authors considered a semiclassical corrections to the production rates due to the interactions between created pairs and suggested a mechanism of colour neutralisation.  

By considering a cylindrical  non-homogeneous  chromoelectic field configurations in \cite{Flory:1983hx}  Flory demonstrated a possible  formation of a chromoelectic flux tube of a finite radius between quark-antiquark pairs  which are  embedded  into the chromomagnetic  gluon condensate (see  the next section).  The alternative mechanisms  of formation of chromoelectric and chromomagnetic flux tubes were considered in \cite{Nielsen:1978nk} and in \cite{Nielsen:1973cs}.

\section{\it Chromomagnetic Gluon Condensate }\label{chromocond}
Let us now apply the renormalisation condition (\ref{renormcondition}) to the Yang-Mills effective  Lagrangian (\ref{effYMlagrangian}) and (\ref{effmagneticlag}). This leads to the following expression for the renormalised effective Lagrangian in chromomagmentic fields \cite{Savvidy:1977as,PhDTheses}:
 \beqa\label{chromomagneticYM}
 \CL^{(1)} &=& { \mu^4 \over 8 \pi^2} \int^{\infty}_{0}  {ds \over s^3} 
\Big( {   a s   \over  ~ \sinh a s } -1- {a^2 s \over 2 }  ({1 \over \sinh s } - {s \cosh s \over \sinh^2 s}) \Big)+\nn\\
&+& { \mu^4 \over 4 \pi^2} \int^{\infty}_{0} {ds \over s^3 }  \Big(   a s   \sin (a s)   - {a^2 s \over 2 }  ({ \sin s } + s \cos s  )  \Big),
 \eeqa
where 
\be
a = g (2 \CF)^{1/2} /\mu^2,    
~~~~~~~ \CF = {1\over 4} G_{\mu\nu}^2  >0 ,~~~
\CG = {1\over 4}  G_{\mu\nu} G^*_{\mu\nu}  =0.
\ee
The Lorentz and gauge invariant field $\CF$ is positive and corresponds to the chromomagnetic field configurations.
The proper time integration can be performed exactly by using the integrals presented in Appendix and  the one-loop SU(2) Lagrangian in terms of Lorentz and gauge invariant  field  $\CF$ is
\be\label{YMeffective1}
\CL^{(1)}   =  - {11  \over 48 \pi^2} g^2 \CF \Big( \ln {2 g^2 \CF \over \mu^4}- 1\Big) 
\ee
and the effective Lagrangian in SU(N) gauge theory  will take the following form \cite{Savvidy:1977as}:
\be\label{YMeffective}
\CL  =  
-\CF - {11  N \over 96 \pi^2}  g^2 \CF \Big( \ln {2 g^2 \CF \over \mu^4}- 1\Big).
\ee
As it follows from this expression, the  QCD vacuum responds to the background chromomagnetic field by inducing a quantum  current of the  charged vector bosons which amplifies the chomomagnetic field imposed on the vacuum. 
The chromomagnetic magnetic induction $\vec{\CB}_a$  of the QCD vacuum  is
\be\label{palarYM}
\vec{\CB}_a = - {\partial \CL \over \partial \vec{\CH}_a} 
= \vec{\CH}_a \Big[1+ {g^2  N \over 96 \pi^2}\log{g^2 \vec{\CH}^2_a \over \mu^4} \Big]~~~= \mu_{vac}~   \vec{\CH}_a .
\ee
The QCD vacuum responds to the background magnetic field as paramagnet  with the magnetic permeability of the following form \cite{PhDTheses}:
\be\label{permeabilityYM}
\mu_{vac} = -{\partial \CL \over \partial \CF}  =1+ {g^2  N \over 96 \pi^2}\log{g^2 \vec{\CH}^2_a \over \mu^4} ~ >~ 1~~~~~\text{\it paramagnetic}.
\ee
The paramagnetism of the QCD vacuum means that it amplifies  the applied chromomagnetic field by generating  induced chromomagnetic field in the direction of the applied field. This phenomenon is similar to the Pauli paramagnetism, an effect associated with the polarisation of the electron spins. In QCD the polarisation of the vector boson spins is responsible for this amplification of the background field (\ref{asymptoticfree}). This also can be seen from the vacuum energy density ( see Fig.\ref{fig2} )
\be\label{energyexpYM}
\epsilon(\CF)=  ~\CF + {11 N \over 96 \pi^2} g^2 \CF \Big( \ln {2 g^2 \CF \over \mu^4}- 1\Big)
\ee
with its new minimum outside of the perturbative vacuum $ < \CF > =0$, at the renormalisation 
group invariant field strength \cite{Savvidy:1977as}
\be\label{chomomagneticcondensate}
 \langle 2  g^2 \CF \rangle_{vac}=    \mu^4  \exp{(-{96 \pi^2 \over 11 N g^2(\mu) })} = \Lambda^4_{QCD}
\ee
characterising the dynamical breaking of conformal symmetry of the  SU(N) gauge field theory\footnote{The  Lorentz invariant  average  $\langle~\rangle$ over the covariantly constant field $G_{\mu\nu}$ orientations can be performed as in \cite{Milshtein:1983th},  \cite{Savvidy:1977as}. In \cite{Milshtein:1983th} the invariant measure was taken in the form $ d\mu =  d \vec{\CH} ~d \vec{\CE} ~\delta(\CF - \CF_{min})~ \delta(\CG^2 - \CG^2_{min})$.  }. The Lorentz invariant form of the effective action (\ref{energyexpYM}) suggests that there are many states which  have the same energy density as the covariantly  constant chromomagnetic field. In a series of articles \cite{Baseyan,Natalia,SavvidyKsystem,SavvidyPlanes} the authors  found and explored spatially  homogeneous solutions of the YM equations which are invariant with respect to the Lorentz transformations and conveniently  represent the gauge field fluctuations in the vacuum.  The average  $\langle ... \rangle$ in (\ref{chomomagneticcondensate}) can  be understood as average over these field configurations (see also \cite{Banks,Anous,Cho:2004qf}).

For the energy momentum tensor (\ref{energymomentum}) we shall get  
\beqa\label{energymomentumYM}
T_{\mu\nu} = T^{YM}_{\mu\nu}\Big[1 +{  11 N  g^2  \over 96 \pi^2} 
  \ln {2 g^2 \CF \over \mu^4} \Big]
- g_{\mu\nu}    { 11  N   \over 96 \pi^2}  g^2 \CF  ,~~~~~~~~~\CG=0.
\eeqa
The trace of the energy momentum tensor is not equal to zero and characterises the breaking of conformal symmetry in  QCD:
\beqa
T  =  T_{\mu\mu}  = -{ 11  N  \over 24 \pi^2} g^2  \CF~.
\eeqa
The vacuum energy density is given in (\ref{energyexpYM}):
$T_{00}=\epsilon(\CF)$ with its minimum  at (\ref{chomomagneticcondensate}) \cite{Savvidy:1977as}.
Substituting this value into the expression for the energy momentum tensor (\ref{energymomentumYM}) we shall get the expression which is proportional to the metric  tensor $g_{\mu\nu} $:
\be\label{vacuumenergyden}
\langle T_{\mu\nu} \rangle_{vac} = - g_{\mu\nu}    { 11  N  \over 96 \pi^2} \langle g^2 \CF \rangle_{vac}
\ee
and is therefore a relativistically invariant characterisation of the vacuum with its negative energy density $\epsilon_{vac}= \langle T_{00} \rangle_{vac}$ and the pressure  $P_{vac}= - \epsilon_{vac}$.
This is an important result because  the vacuum state should be Lorentz invariant and  its stress tensor $T_{\mu\nu}$ should be the same in all frames \cite{Zeldovich:1968ehl,Weinberg}.  As a result, its vacuum average value can only be of the cosmological type $\langle  T_{\mu\nu}  \rangle = \epsilon_{vac}\ g_{\mu\nu}$, and indeed it is.

Let us consider the  behaviour of the effective Lagrangian from the renormalisation group point of view and compare it with the behaviour  of the effective coupling constant.
The equations derived above are universally true for the non-Abelian field as well. Thus when  $\CG = \vec{\CE_a} \vec{\CH_a} =0 $ we have  
\be\label{effectivecoupling}
\CM(t,g) = {\partial \CL \over \partial \CF}= - {g^2 \over \bar{g}^2(t)},~~~~~~{d \bar{g} \over dt } = \bar{\beta}(\bar{g}) .
\ee
The   vacuum magnetic permeability  introduced in (\ref{permeability})  will take the following form  \cite{PhDTheses}:
\be\label{permeabilityfulYMl}
{\it \mu}_{vac}~ = ~ {g^2 \over \bar{g}^2(t)}  ,~~~~~~~~~\CG=0. 
\ee
The Callan-Symanzik beta function can be calculated by using (\ref{YMeffective}): 
\be\label{betafunction}
\bar{\beta}=  {1\over 2}g {\partial \CM \over \partial t} \vert_{t =0}= -{11  N \over 96 \pi^2}g^3
\ee
and the effective coupling constant as a function of the field has the form
\be\label{effectivecouplingYM}
\bar{g}^2(\CF) = {g^2  \over 1+ {  11 g^2  N \over 96 \pi^2} 
  \ln{2 g^2 \CF \over \mu^4}  }, 
\ee
where we introduced the Casimir operator $C_2(G)=N$ for the gauge group $G=SU(N)$.

Let us consider the value of the field strength tensor $\CF_{0}$ at which the vacuum energy density (\ref{energyexpYM}) vanishes  $\epsilon(\CF_0)=0$, as it is  shown on  Fig.\ref{fig2}:
\be\label{intersectionspoint}
2 g^2 \CF_{0}=   \mu^4 \exp{(-{96 \pi^2 \over 11 g^2 N} +1)} = e  \langle 2  g^2 \CF \rangle_{vac}.
\ee 
The effective coupling constant (\ref{effectivecouplingYM}) at this field strength has the value 
\be\label{intersection}
\bar{g}^2(\CF_0) = {96 \pi^2   \over   11 N}~. 
\ee
It follows  that the effective coupling constant at the intersection point $\CF_0$ is small: \be\label{effcoupling}
\bar{g}^2(\CF_0) = {96 \pi^2   \over   11 N}   \ll 1 ~~~\text{if}~~~ N \gg   {96 \pi^2   \over   11} ~.
\ee
The energy density curve $\epsilon(\CF)$ (\ref{energyexpYM}) intersects the horizontal zero energy line  at the nonzero angle  $\theta$ (see Fig.\ref{fig2}):
\be\label{intersectionangle}
\tan\theta={ 11 g^2 N \over  96 \pi^2 }  ~>~ 0.
\ee
This means that  i) the true vacuum state is  below the perturbative vacuum  and that ii) there is a nonzero chromomagnetic field in the vacuum.  Now the question is, how far into the infrared region one can continue the energy density curve by using the perturbative result?  Let us consider the fields which are approaching the infrared pole. This can be done, in particular, by using the following parametrisation: 
\be\label{infraredpole}
  \CF_{n}=e^{1-n} \langle  \CF \rangle_{vac},
\ee
where the parameter $n$ is less than one, and we have  $ \CF_{n} ~\rightarrow  ~ \langle  \CF \rangle_{vac}$ when $n$ tends to unity from below.  At these fields values the effective coupling constant (\ref{effectivecouplingYM})  tends to zero:
\be\label{infraredpole1}
\bar{g}^2(\CF_n) = {96 \pi^2   \over   11 N (1-n)} \rightarrow  0
\ee
if the product  $N (1-n) \rightarrow \infty$ is large and the t'Hooft coupling constant $g^2 N = \lambda$ is fixed and small.  It follows then that the effective coupling constant can be made small to justify  the use of the perturbative result and  the energy density curve can be continuously  extended infinitesimally close to the value of the vacuum field $\langle  \CF \rangle_{vac}$, as it is shown on Fig. \ref{fig2}.    Let us  analyse how the field at the intersection point (\ref{intersectionspoint}) and the effective coupling constant  (\ref{intersection}) are changing when we include the two-loop contribution. The two-loop\footnote{The beta   function (\ref{effectivecoupling}) coefficients  $\bar{\beta} = \beta_1 g^3 + \beta_2 g^5 +..$  are  given by $\beta_1 =- {11 N   \over 6 (4\pi)^2 } $ and $\beta_2 = -{34  N^2 \over 6 (4 \pi)^4 } $ \cite{Jones:1974mm,Caswell:1974gg}. } effective Lagrangian has the form \cite{PhDTheses}
\be
\CL  =  
-\CF - \Big({11   \over 6 (4\pi)^2 } g^2 N + {34   \over 6 (4 \pi)^4 } (g^2 N)^2 \Big)  \CF \Big( \ln {2 g^2 \CF \over \mu^4}- 1\Big).   
\ee
The field at the intersection point (\ref{intersectionspoint}) is shifted  by an exponentially small correction
\be\label{effectivecouplingYM1}
2 g^2 \CF'_{0}=  \exp{\Big(-{96 \pi^2 \over 11 \lambda} \cdot{1\over 1 + {17 \over 88 \pi^2} \lambda } +1\Big)}.
\ee
At this field the effective coupling constant is smaller by the factor $1/  1 + {17 \over 88 \pi^2} \lambda  $ 
\be\label{intersection1}
\bar{g}^2(\CF'_0)  ={96 \pi^2   \over   11 N} \cdot {1   \over   1 + {17 \over 88 \pi^2} \lambda  } \ll 1,
\ee
and  the inequality (\ref{intersection1}) is fulfilled at smaller values of $N$ than in the first approximation (\ref{effcoupling}).   The  chromomagnetic  condensate in the two-loop approximation will take the following form:   
\be\label{twoloopcond}
\langle 2  g^2 \CF \rangle_{vac}=  
 \mu^4  \exp{\Big(~  -{1\over  \beta_1 g^2} \Big[1 - {\beta_2 g^2 \over \beta_1} \ln(1 +   {\beta_1 \over \beta_2 g^2} )\Big]  ~       \Big)}.
\ee
The high-loop corrections are analysed below  by using renormalisation group results (\ref{effectivecouplingel}), (\ref{effectivecoupling}) and the expression (\ref{twoloopcond})  can be recovered through the general expression (\ref{chomomagneticcondensaterenormgroup}).

It is interesting  to know if  the energy density curve is a continuous function of the field strength $\CF$ in the region $[0, \langle \CF \rangle_{vac}]$ which is outside of  
the validity of the perturbative calculations and if  the energy density curve is a convex function.   
A non-perturbative functional method developed by Zwanziger in \cite{Zwanziger:1982na}  is answering to these questions affirmatively.  It seems  that further development of his approach can shed even more light to the behaviour of the effective Lagrangian in the nonpertubative region.  We already obtained  the first derivative  of the energy density curve (\ref{effectivecoupling}) and can calculate its second derivative as well:
\beqa\label{derivative2} 
{\partial \epsilon \over \partial \CF} =  - {\partial \CL \over \partial \CF}  = {g^2 \over \bar{g}^2}, ~~~~~~~~~~
  \CF {\partial^2 \epsilon \over \partial \CF^2}=    {\partial   \over \partial t} {g^2 \over \bar{g}^2(t)} = -g^2  ~{\bar{\beta}(\bar{g})  \over \bar{g}^3}.
\eeqa
The sign of the second derivative depends on the sign of the  beta function. In QCD,  in the perturbative regime this ratio is negative and the second derivative (\ref{derivative2}) is positive:
\be
\CF {\partial^2 \epsilon \over \partial \CF^2}= {11   \over 6 (4\pi)^2 } g^2 N + {34   \over 6 (4 \pi)^4 } (g^2 N)^2 +.... 
\ee
Thus the energy density curve is convex (see Fig.\ref{fig2}).  In QED the overall sign is  negative and the energy density curve is concave (see Fig.\ref{fig1}). 

{\it Any non-perturbative information about the ratio $\bar{\beta}(\bar{g})/\bar{g}^3$ can be translated into the information about property of the energy density curve $\epsilon(\CF)$. 
As far as the beta function $\bar{\beta}(g)$ has no zeros,  is a negative  analytical function of the coupling constant  and 
$\int^{\infty}_{g} {d g \over \bar{\beta}(g)} < \infty$, 
then it follows from (\ref{derivative2}) that the energy density curve is convex  and that the minimum of the energy density curve $\epsilon(\CF)$ is defined by the extremum where its first derivative vanishes.} Using the expression (\ref{effectivecoupling}) one can derive the value of the chromomagnetic condensate   \cite{Savvidy:1977as}:
\be\label{chomomagneticcondensaterenormgroup}
\langle 2  g^2 \CF \rangle_{vac}=    \mu^4  \exp{\Big(2 \int^{\infty}_{g}{d g \over \bar{\beta}(g)}\Big)}.
\ee
To all orders in the perturbation theory the derivative of the energy momentum tensor trace can be obtained by using the renormalisation group invariant result (\ref{renormtraceenergy}):  
\be
{\partial T \over \partial \CF} =  4  g^2 ~{\bar{\beta}(\bar{g}(t))  \over \bar{g}(t)^3}~, ~~~~~t = {1\over 2}\ln(2 g^2 \CF/ \mu^4)~.
\ee
Integration over $\CF$ gives the trace
\be
T_{\mu\mu} = 4 \int { g^2 \over \bar{g}^2 }  ~{\bar{\beta}(\bar{g})  \over \bar{g}} d \CF.
\ee
If one considers the approximation in which the effective coupling constant (\ref{effectivecouplingel}) is field independent $\bar{g}(t) \equiv g$ then this formula after  integration over $\CF$ will reduce to the one given in literature \cite{Adler,Nielsen,Minkowski}:
 \be
 T_{\mu\mu}  = 4 {\bar{\beta}( g)  \over  g} \CF.
 \ee
Otherwise the field dependence of the energy momentum trace is defined through the beta functions and effective coupling constant and has more complicated dependence on field strength tensor $\CF$.

\section{\it Effective Cosmological Constant}

As is follows from  (\ref{vacuumenergyden}), in the ground state the following relation between energy density $\epsilon_{vac}$ and pressure  $P_{vac}$ takes place:
$ 
\epsilon_{vac} = - P_{vac}  
$. It is a relativistically invariant characterisation of the vacuum \cite{Zeldovich:1968ehl,Weinberg}, and it represents  a field-theoretical  contribution  into the effective cosmological constant  $ \Lambda_{eff}$:
\be
\epsilon_{vac}  = {c^4 \over 8 \pi G  }  \Lambda_{eff}~=-{ b    \over 192 \pi^2}  \langle 2 g^2 \CF \rangle_{vac}~=-{ b    \over 192 \pi^2}  \Lambda^4_{QCD} . 
\ee
The chromomagnetic condensate (\ref{chomomagneticcondensate}) is of order $\Lambda^4_{QCD}$, and the vacuum energy density is negative and is about $\epsilon_{vac} \approx -b\ 10^{-8}  GeV^4 $ . The value of the cosmological constant measured in the observation of the high-z Type Ia supernovae  \cite{Riess:1998cb,Tonry:2003zg,Perlmutter:1998np,Betoule:2014frx} and by the Plank Collaboration  \cite{Adam:2015rua,Aghanim:2018eyx}
 $
 \epsilon_{\Lambda}= c^4  \Lambda_{obser} /   8 \pi G   \approx 10^{-47} GeV^4
 $
is  about 39 decimal places smaller and positive. It is important to mention that the energy gap depends on a gauge group and a matter content, the beta function (the $b$ parameter in one-loop approximation), as well as of the temperature of the universe \cite{Kay1983}. At high temperatures the curve of the effective potential moves upward, the value of the chromomagnetic gluon condensate tends to zero, as well as the $\Lambda_{eff}$, and the scaling invariance get restored. The phase transition is of the second-order  \cite{Kay1983}. 

In the article \cite{Pasechnik:2016twe} the authors suggested a possible cancelation mechanism between chromomagnetic and its "mirror chromoelectric"  condensates. In this proposal, which involves adding to the SM particles a mirror world (dark matter) 
\cite{Oort,Zwicky,Nishijima,Foot:2003eq,Berezhiani:2003xm,Barbieri:2005ri},  the entire SM is replicated in a mirror world.  The new $Z_2$ symmetry interchanges SM with the mirror SM, ensuring identical particles and interactions. It is conjectured  that the quantum vacua of the  "Mirror SM"  contribute to the cosmological constant on the same footing as the SM, since mirror particles are expected to gravitate in the same way as the usual ones,  and that the mirror chromoelectric  gluon condensate contributes to the energy density of the universe with a positive sign and thus may, in principle, eliminates the negative QCD vacuum effect by yielding a cosmological constant small. The alternative mechanism was considered in \cite{Addazi:2019mlo}.

In electroweak theory  the Higgs vacuum field also generates a negative contribution to the effective cosmological constant  \cite{Linde:1974at,Linde:1980ts}.  In both field theories the value of the $\Lambda_{eff}$ is of many orders of magnitude larger than the observed value.  
 
\section{\it Absence of Imaginary Part in Chromomagnet Field}
A careful inspection of the charged vector bosons spectrum in a chromomagnetic field by  Nielsen, Olesen  \cite{Nielsen:1978rm} and Skalozub \cite{Skalozub:1978fy} demonstrated  that due to the unstable (higgs-like) mode    
$
k^2_0 = k_{\vert \vert}^2 - g f_1  
$
~($k_{\vert \vert}^2 \leq g f_1$) there is an imaginary part in the effective Lagrangian: 
 \beqa
Im \CL^{(1)} = Im ~ {g f_1 \over 4 \pi^2}~ \int^{\infty}_{- \infty} d k_{\vert \vert}  \sqrt{k_{\vert \vert}^2 - g f_1 -i \epsilon} = -  {g^2 f^2_1  \over 8 \pi}.  \nn
\eeqa
In the subsequent publications \cite{Nielsen:1978zg,Ambjorn:1978ff,Nielsen:1978tr,Nielsen:1979xu,Nielsen:1979ta,Ambjorn:1979xi,Ambjorn:1980ms,Zwanziger:1982na,Flory:1983td,Kay1983,Cho:2004qf,Consoli:1985sx} the theoretical groups at the Niels Bohr Institute,  New York University, SLAC  and  Bari University came to the conclusion that due to the quartic self-interaction term in the YM action there is a hidden  Higgs mechanism, which stabilises the covariantly constant  field configurations  so that  the effective Lagrangian remains a real function in the  background chromomagnetic  field \cite{Ambjorn:1978ff,Nielsen:1979xu,Ambjorn:1980ms,Zwanziger:1982na,Flory:1983td}. This is the reflection of the fact that calculations are performed in the approximation in which only the quadratic term  $\delta A_{\mu} H_{\mu\nu} \delta A_{\nu}$ of the fluctuating fields $\delta A_{\mu}$  in the direction of the  unstable mode $\Phi $  was taken into the consideration in the loop expansion (\ref{loopexpansion}) and  (\ref{oneloopeffec}). The quartic term $V_{\mu\nu\lambda\rho} \delta A_{\mu}  \delta A_{\nu}  \delta A_{\lambda}  \delta A_{\rho}$ of the Yang-Mills Lagrangian should be taken into consideration in this circumstance and play a crucial role in stabilising the quantum mechanical fluctuations \cite{Nielsen:1978nk,Nielsen:1979xu,Ambjorn:1978ff,Ambjorn:1980ms,Flory:1983td}. The phenomenon is similar to the one in the classical Higgs Lagrangian where the zero field configuration is unstable and a field configuration at the bottom of the potential provides a stable field configuration due to the quartic term.  In the pure YM theory the higgs-like action for the unstable mode  was derived in \cite{Nielsen:1978nk,Ambjorn:1978ff,Nielsen:1979xu,Ambjorn:1980ms}. It was proposed to  search the stable solutions of the classical Yang-Mills field equations in a fixed background chromo-magnetic field which plays the role of an external order parameter.  Without the presence of the order mass parameter, the $g H$ in the given case,   the conformal invariance of the pure classical Yang-Mills equations prevents the existence of localised solutions \cite{Nielsen:1973cs}.  This program was successfully realised  with the discovery of the field configurations which are varying in space due to the development of the unstable mode, the colour magnetic flux tubes and  the spaghetti magnetic tubes forming the domain-like field configurations \cite{Nielsen:1979xu}. The configurations are supported by the external chromomagnetic field $gH$. The difficulty of this approach lies in the calculation of  the quantum-mechanical fluctuations around these  field configurations and to see that the solutions remain localised when the external field is switched off.  The important conclusion of the investigation was that it pointed out to the fact that the stability of the chromomagnetic field configurations is a natural consequence of the quartic self interaction of the unstable mode\footnote{ A non-perturbative prove of the reality and concavity of the effective action is due to Zwanziger \cite{Zwanziger:1982na}. }.

This result became the initial point for the investigation initiated by Curt Flory in his  article devoted to the resolution of the higgs-like mode problem \cite{Flory:1983td}. His breakthrough idea was to integrate exactly  the  functional integral over the higgs-like mode from the start in order to get the  quantum-mechanical contribution to the effective Lagrangian corresponding to the summation of the contributions  coming from all loop diagrams  with higgs-like mode propagation in the loops. 
Presenting the amplitude of the higgs-like mode and of the corresponding action in terms of dimensionless variables $k_{\mu} \rightarrow k_{\mu}/ \sqrt{gH}$, $x_{\mu} \rightarrow x_{\mu}  \sqrt{gH}$ one can get\footnote{In the SU(2) case $W_{\mu} = {1\over \sqrt{2}} (A^1_{\mu} + A^2_{\mu})$, $A_{\mu}=A^3_{\mu}$ and $W = W_1 = -i W_2$, as it is defined in \cite{Nielsen:1978nk,Ambjorn:1978ff,Nielsen:1979xu,Ambjorn:1980ms}. }:  
\be\label{unstablemode}
(gH)^{-1/2} ~W =  \int {d k_2 \over 2 \pi} e^{-{1\over 2}(x_1 + k_2 )^2}  \Phi_{k_{2}}(x_0,x_3),
\ee
where $\Phi_{k_{2}}(x_0,x_3)$ is the dimensionless amplitude of the higgs-like mode.  The action of the higgs-like mode \cite{Ambjorn:1978ff} will take the following form: 
\be\label{unstablelagrangian}
{S_{higgs\ mode} \over \sqrt{2 \pi}} =  \int {d k_{2} \over 2 \pi } d x_0 d x_3 \Big( \vert \partial_{\mu} \Phi_{k_{2}} \vert^2 +  \vert  \Phi_{k_{2}} \vert^2 
- {1\over 2} g^2  \int {d p d q \over (2 \pi)^2 }  e^{-{p^2 +q^2\over 2}} \Phi^*_{k_2 +p} \Phi^*_{k_2 +q}\Phi_{k_2} \Phi_{k_2 +p+q}\Big).
\ee
The  Feynman integral over the higgs-like mode $ \Phi$ with the above action will corresponds to the summation of all loop diagrams with the higgs-like mode propagation in the loops. The beauty of this approach lies in the fact that this integral can be computed exactly. Indeed, 
what is essential in this representation is that the dependence on the chromomagnetic field does not show up in the action (\ref{unstablelagrangian}) and appears only in front of the higgs-like field amplitude $(gH)^{1/2}$ in (\ref{unstablemode}). Therefore  the integration of the action (\ref{unstablelagrangian}) over  the field $\Phi_{k_{2}}$ is background field independent. Thus the contribution of the higgs-like mode to the effective Lagrangian is only through the integration measure and its degeneracy: 
$$
e^{S_{higgs~mode}}  \approx (C g H)^{-{1\over 2}  ({gH \over 2\pi})^2 } = \exp{\Big(- {g^2 H^2 \over 8 \pi^2}( \log g H +\log C~)~\Big)},
$$ 
where $C$ is the $gH$  independent value of the functional integral over $\Phi_{k_{2}}(x_0,x_3)$.  This contribution is a real function of chromomagnetic field \cite{Flory:1983td,Kay1983}.  After taking into account the contributions from all other positive modes the effective Lagrangian takes the form which identically coincides with  (\ref{YMeffective1}). This confirms the expression (\ref{YMeffective1}) being without imaginary part (\ref{imaginarypart}).

One can consider the above approach of calculating the effective action as an alternative to a standard loop expansion in the following sense:  The expansion is organised by rearranging the perturbative expansion (\ref{loopexpansion}) in a background field $A$ so that the quartic self-interactions of eigenmodes are  included into the propagator of the gauge field $G(x,y;A) $ and  the   loop expansion is performed in terms of the  remaining  cubic and cross-mode quartic vertices of the YM action. 

\section{\it Conclusion}

The short overview of the publications devoted to the chromomagnetic gluon condensation and QCD vacuum are given below.  The confinement problem from the point of view of the QCD vacuum and chromomagnetic gluon condensate was considered in the articles of Mandelshatam \cite{Mandelstam:1979xd,Mandelstam:1979ir,Mandelstam:1980ii}, Nambu \cite{Nambu:1981gt},  Adler and Piran \cite{Adler:1983zh} and  Nielsen and Olesen \cite{Nielsen:1979vb}.   The thermodynamics of the Yang-Mills gas by Linde \cite{Linde:1980ts}. The publication on generation of galactic magnetic field due to the condensation of vector field was considered in  \cite{Enqvist:1994rm}.  The induced gravity was considered by Adler \cite{Adler:1982ri}. The  magnetostatics  was considered in  \cite{Pagels:1978dd}. The phenomenology of hadrons and the properties of the QCD vacuum by Shuryak  \cite{Shuryak:1984nq}.  The mechanism of dynamical supersymmetry breaking and string compactification  to four dimension due to the properties of the non-Abelian effective action was suggested by Veneziano and Taylor \cite{Taylor:1988vt}. The dynamical mass generation in QCD and glueballs spectrum by Cornwall  \cite{Cornwall:1981zr,Cornwall:1983zb}. The string-like solution of pure YM equations stabilised in the presence of the chromomagnetic condensate by Faddeev and Niemi \cite{Faddeev:2001dda,Faddeev:2006bm} and the monopole condensation by Cho \cite{CHO:2014zza}. A  complementary to Zwanziger \cite{Zwanziger:1982na} a non-perturbative approach for the construction of effective actions at different scales, the Wilsonian effective actions,   was developed by Reuter and Wetterich in series of articles \cite{Reuter:1994zn,Reuter:1994yq,Reuter:1993kw,Wetterich:1992yh}. 

Alternative  approach for the investigation of the condensates in Yang-Mills theory is provided by the Monte Carlo  lattice simulations \cite{DiGiacomo:1981lcx,Kripfganz:1981ri,DiGiacomo:1981dp,Ilgenfritz:1982yx,Bali:2014sja}.  The lattice formulation  is offering a non-perturbative regularisation of the YM theory and in principle allows to measure different QCD condensates. One of the aims of these calculations was to extract a non-perturbative value of the vacuum expectation value (VEV) of the  composite operator $G^2_{\mu\nu}$ :   
\be\label{actionvev}
\langle 0 \vert  {\alpha_s \over \pi} G_{\mu\nu} G_{\mu\nu} \vert 0 \rangle.
\ee
In perturbation theory this  VEV is diverging as the fourth power of the cutoff and after renormalisation is set to zero.  In our investigation we were studying a different observable, the effective action $\Gamma[A]$ (\ref{effectiveaction}), which depends on the VEV of the gauge field operator $ \langle 0\vert A^{a}_{\mu}(x) \vert 0 \rangle  \equiv  A^{a}_{\mu}(x) $.

As it was stressed in references   \cite{DiGiacomo:1981lcx,Kripfganz:1981ri,DiGiacomo:1981dp,Ilgenfritz:1982yx,Bali:2014sja}, the main difficulty in measuring the condensates of the type (\ref{actionvev}) lies in the necessity to subtract from the Monte Carlo data the dominant (and diverging)  perturbative  contribution  and then to extract the exponentially falling non-perturbative term of the form (6.99), which is the only one of interest from the point of view of the continuum theory. The evaluation of the VEV  requires the calculation of the following expression:  
\be
\lim_{a \rightarrow 0}~  {C\over a^4}~  \langle 0 \vert  (1-P) \vert 0 \rangle_{meas} - \langle 0 \vert  (1-P) \vert 0 \rangle_{pert} =   \langle 0 \vert  {\alpha_s \over \pi} G_{\mu\nu} G_{\mu\nu} \vert 0 \rangle +...,
\ee
where $P$ is  the plaquette operator and dots denote the operators of higher dimension. The perturbative VEV  is represented by a diverging series  \cite{Dyson:1952tj,Lipatov:1976ny,tHooft}:
\be\label{pert}
\langle 0 \vert  (1-P) \vert 0 \rangle_{pert} = \sum^{\infty}_{n=0} c_n \alpha^n_s .
\ee 
 As it was stressed in  \cite{DiGiacomo:1981lcx,Kripfganz:1981ri,DiGiacomo:1981dp,Ilgenfritz:1982yx,Bali:2014sja}, the separation of perturbative and non-perturbative contributions has  arbitrariness which makes any determination of  the QCD condensates in terms of composite operator (\ref{actionvev}) ambiguous.       
 
The  phenomena of chromomagnetic gluon condensation \cite{Savvidy:1977as} initiated series of publications by the ITEP group where they used the gluon condensate to improve their perturbative sum rule equations \cite{Novikov:1976tn,Zakharov:1999jj} \footnote{In 1977 the author gave a theoretical seminar on the chromomagnetic gluon condensation \cite{Savvidy:1977as} in ITEP.    At end of the seminar  one of the participants, Victor Novikov, on our way back to the metro station by  tram, remarked  to the author  that the theoretical prediction of the chromomagnetic  condensate presented at the seminar \cite{Savvidy:1977as} can  be crucial in improving  the naive sum rule equations  published earlier in \cite{Novikov:1976tn}  by introducing  the chromomagnetic condensate in the form of power corrections.    A year later, the proposal  was  realised  in \cite{Zakharov:1999jj}.  }. Modern determination of the gluon condensate numerical value from hadronic $\tau$ decay data and from the charmonium sum rules  can be found in the review article of Ioffe \cite{Ioffe:2010zz}. The best values of condensates, extracted from QCD sum rules from experimental data, are given in Table 1 in \cite{Ioffe:2010zz}. These data do not allow to exclude the zero value for the gluon condensate \cite{Zyablyuk:2002kg,Samsonov:2004zm,Ioffe:2010zz}.  The separation of perturbative and non-perturbative contributions has  arbitrariness  \cite{Zakharov:1999jj} (similar to the lattice calculations), as it was pointed out by Ioffe in \cite{Ioffe:2010zz}.   

Here we reexamined the phenomena of the YM condensation investigating the effective action $\Gamma[ A ]$ (\ref{effectiveaction}). It is of the chromomagnetic type and it has a numerical value  $\Lambda^4_{QCD}$ which is of the order of few hundred $MeV^4$ 
\be
 \langle 2  g^2 \CF \rangle_{vac}=  \langle {g^2 \over 2} G^2_{\mu\nu} \rangle_{vac} = \langle  g^2 (\vec{\CH}^2_a - \vec{\CE}^2_a ) \rangle_{vac} =   \mu^4  \exp{\Big(2 \int^{\infty}_{g}{d g \over \bar{\beta}(g)}\Big)} = \Lambda^4_{QCD}~  >~ 0
 \ee 
or in terms of the strong coupling constant 
\be
\langle {\alpha_s \over \pi} G^2_{\mu\nu} \rangle_{vac} = \langle {g^2 \over 4 \pi^2} G^2_{\mu\nu} \rangle_{vac}= 
{\Lambda^4_{QCD} \over 2 \pi^2}~ . 
 \ee

\section{\it Appendix A} 
The integrals appearing in the effective Lagrangian  (\ref{masslessHE}) have the following form:
\beqa\label{integral}
&&\int^{\infty}_{0}  
  {   ds \over  s^{1-k} ~ \sinh^2( a s) }= {4  \over (2 a)^k}  \Gamma(k)  \zeta(k-1), \\ 
&& \int^{\infty}_{0}  
 {    \cosh(b s) ds \over  s^{1-k}~ \sinh( a s) }= {\Gamma(k) \over (2 a)^k} \Big[ \zeta(k, {1\over 2}(1- {b \over a}) + \zeta(k, {1\over 2}(1+ {b \over a})\Big], ~~~b \neq a,\nn
\eeqa
where  $k$ can be considered as a dimensional regularisation parameter  and the integrals should be calculated in the limit $k \rightarrow -1$ \cite{PhDTheses}.  As far as $b \neq a$,  the second integral does not completely coincide with the one appearing in the effective Lagrangian (\ref{masslessHE}) and we have to consider its extension. In order to calculate  the integral we shall take $b=a-\epsilon$  and consider the  limit  $\epsilon \rightarrow 0$:
\beqa
\int^{\infty}_{0}    { s^{k-1}    \cosh((a-\epsilon) s) ds \over  ~ \sinh( a s) } &=& {\Gamma(k) \over (2 a)^k} \Big[ \zeta(k, {\epsilon \over 2 a} ) + \zeta(k, (1- {\epsilon  \over 2 a})\Big].\nn
\eeqa
By the definition the Riemann zeta function $\zeta(k, q)$ is \cite{gradshteyn}
\be
\zeta(k, q) =\sum^{\infty}_{n=0}{1\over (n+q)^k}= \sum^{\infty}_{n=1}{1\over (n+q)^k}+ {1\over q^k}\nn
\ee 
and in the limit $q=\epsilon/2 a  \rightarrow 0$ we shall get
\beqa
  \lim_{\epsilon   \rightarrow 0}  \zeta(k, {\epsilon \over 2 a} )     ~= ~ \zeta(k) +  ({2 a \over \epsilon})^k,~~~~~~~~~
   \lim_{\epsilon   \rightarrow 0} \zeta(k, (1- {\epsilon  \over 2 a})~= ~ \zeta(k). \nn
\eeqa 
Thus the second integral in (\ref{integral}) takes the following form: 
\beqa
\int^{\infty}_{0}    { s^{k-1}    \cosh((a-\epsilon) s) ds \over  ~ \sinh( a s) } 
={\Gamma(k) \over (2 a)^k} \Big[ 2 \zeta(k)   +  ({2 a \over \epsilon})^k\Big]=
 {2 \Gamma(k)  \zeta(k) \over (2 a)^k}  +  {  \Gamma(k) \over \epsilon^k}, \nn
\eeqa
where the last term is field independent and can be subtracted from the effective Lagrangian.

The integrals appearing in Yang-Mills effective Lagrangian (\ref{chromomagneticYM}) have the form  \cite{PhDTheses}
\beqa
&&\int^{\infty}_{0}  
  {   ds \over  s^{1-k} ~ \sinh( a s) }= {2^k -1  \over 2^{k-1} a^k}  \Gamma(k)  \zeta(k), ~~~~
 \int^{\infty}_{0}  
 {    \cosh(a s) ds \over  s^{1-k}~ \sinh^2( a s) }= {2^{k-1} -1  \over 2^{k-2} a^k}  \Gamma(k)  \zeta(k-1) \nn\\ 
&& \int^{\infty}_{0}  
 {    \sin(a s) ds \over  s^{1-k}}=  {\Gamma(k) \over a^k} \sin{k \pi \over 2},~~~~~~~~~~~~~~
\int^{\infty}_{0}  
 {    \cos(a s) ds \over  s^{1-k}}= {\Gamma(k) \over a^k} \cos{k \pi \over 2} .\nn
\eeqa
The Lagrangian should be calculated in the limit $k \rightarrow -1$.

\section{\it Appendix B}   

As it follows from (\ref{energymomentum})- (\ref{renormtraceenergy}) in the pure chromomagnetic  case  we have the relations (\ref{derivative2})
$$
{\partial \epsilon \over \partial \CF} = {g^2 \over \bar{g}^2},~~~~~~~~  \CF {\partial^2 \epsilon \over \partial \CF^2}=     -g^2  ~{\bar{\beta}(\bar{g})  \over \bar{g}^3},~~~~~~~~{d \bar{g} \over dt } = \bar{\beta}(\bar{g}),
$$
which define the behaviour of the energy density curve $\epsilon(\CF)$ in terms of the beta function.

\end{document}